\documentclass[11pt,a4paper]{article}
\pdfoutput=1
\usepackage{jheppub}
\usepackage{graphicx}
\usepackage{comment}
\usepackage{bm,array}
\usepackage{graphics}
\usepackage{epsf}
\usepackage{color}
\usepackage{amsmath}
\usepackage{amssymb}
\usepackage{latexsym}
\usepackage{slashed}
\usepackage{float}
\usepackage{cases}
\usepackage{multirow}
\def\ba{\begin{array}}
\def\ea{\end{array}}
\def\l{\lambda}

\def\wplus{W^+}
\def\wminus{W^-}

\def\stone{\tilde{t}_1}
\def\sttwo{\tilde{t}_2}
\def\sbone{\tilde{b}_1}
\def\stones{{\tilde{t}}^*_1}
\def\sttwos{{\tilde{t}}^*_2}

\def\ntrlone{\tilde{\chi}_1^0}

\def\charonepm{\tilde{\chi}_1^\pm}
\def\mstone{m_{\tilde{t}_1}}
\def\msttwo{m_{\tilde{t}_2}}
\def\msbone{m_{\tilde{b}_1}}
\def\mntrlone{m_{\tilde{\chi}_1^0}}

\def\mcharonepm{m_{\tilde{\chi}_1^\pm}}
\newcommand{\fbinv}{\,\text{fb}^{-1}}
\newcommand{\beq}{\begin{equation}}
\newcommand{\eeq}{\end{equation}}
\newcommand{\bea}{\begin{eqnarray}}
\newcommand{\eea}{\end{eqnarray}}
\title{Two Light Stops in the NMSSM and the LHC}
\author[a, b]{Jyotiranjan Beuria}
\author[a,b]{Arindam Chatterjee}
\author[a]{AseshKrishna Datta}
\author[a,b]{Santosh Kumar Rai}

\affiliation[a]{Harish-Chandra Research Institute, Allahabad 211019, India}
\affiliation[b]{Regional Centre for Accelerator-based Particle Physics \\
                Harish-Chandra Research Institute, Allahabad 211019, India} 

\emailAdd{jyotiranjan@hri.res.in, arindam@hri.res.in, asesh@hri.res.in, 
skrai@hri.res.in}
\preprint{HRI-P-15-05-001 \\ 
\vspace*{-0.8cm}
\begin{flushright}
RECAPP-HRI-2015-007
\end{flushright}
}
\abstract{We study the viability of having two relatively light top squarks 
(`stops') in the framework of the Next-to-Minimal Supersymmetric Standard Model 
(NMSSM). Such light stops render the NMSSM rather `natural'. These are shown to 
be allowed by the relevant direct searches at the Large Hadron Collider (LHC) 
and to be compatible with the latest LHC results on the Higgs sector, other low 
energy electroweak constraints and recent constraints from the dark matter (DM) 
sector. We propose dedicated searches for such light stops at the LHC within a 
`simplified' scenario that may have a bino-like or a singlino-like neutralino
LSP as the DM candidate and point out various final states carrying the imprint 
of their collective presence. Under certain circumstances, in such a scenario, 
presence of two light stops may give rise to final states which are not so 
typical in their search.  Thorough studies at the detector level reveal the 
status of such a scenario after the 8 TeV run of the LHC and shed light on the 
prospects of its 13 and 14 TeV runs. In favorable regions of the NMSSM parameter 
space, with low-lying spectra, signals with significance $\gtrsim 5\sigma$ are 
possible with a few tens to a few hundreds of $\fbinv$ of integrated luminosity 
in diverse final states.} 
\begin{document} 
\maketitle
%
\section{Introduction}
\label{sec:intro}
The observation of a scalar boson, now appearing more and more to be the Higgs 
boson, by the ATLAS and the CMS collaborations \cite{Aad:2012tfa,Chatrchyan:2012ufa} 
of the Large Hadron Collider (LHC), has accounted for the last missing piece in
the particle spectrum of the Standard Model (SM) of particle physics and has 
vindicated the mechanism of the breaking of the electroweak symmetry as 
incorporated in the SM. Notwithstanding the fact that the SM can now safely be 
considered as an essential part of our understanding of the micro-world, there 
are ample reasons to believe that it does not offer the complete picture of what 
Nature has chosen for us. There are issues which are not understood by staying 
within the SM, and require ideas extending the SM-picture of particle 
physics. This has led to a plethora of models beyond the SM (BSM) which one way 
or the other predict new exotic particle states in addition to the SM ones.
The experimental results from the ATLAS and CMS experiments with center of mass 
energies of 7 and 8 TeV, have not only confirmed the predictions of the SM, 
including that of the Higgs boson, but have also started pushing the 
energy scale up for a possible BSM physics scenario. 

Supersymmetry (SUSY) has been one of the most popular BSM scenarios and is 
rather thoroughly studied in the last three decades. Its simplest incarnation, 
the so-called minimal supersymmetric standard model (MSSM) offers a rich 
phenomenology involving the SUSY particles (sparticles, the partners of the SM 
excitations differing only in their spins), that would be present below the 
$\mathcal{O}$(TeV) scale on the ground of naturalness and hence, supposed to be 
within the reach of the LHC experiments.  However, no hint of such excitations 
has been found in the experiments to date and data from the LHC have already 
put rather stringent limits on the masses of such particles \cite{susylimits}.  
In addition, the recent observation of a scalar resonance at $\sim 125$ GeV 
followed by the studies of its properties reveal an impressive agreement with 
the predictions of the SM. These have put the MSSM under tremendous scrutiny. 
Although a light CP-even scalar eigenstate with a mass of 125 GeV is possible 
in the MSSM through radiative corrections, it prefers the third generation squark 
mass (in particular, the stop mass) to be much larger than $\mathcal{O}$(1 TeV) 
and therefore puts the most natural solution to the so-called ``gauge hierarchy 
problem'' on somewhat troubled grounds by requiring an uncomfortable degree of 
fine-tuning. 

Thus, within the MSSM, on the ground of naturalness, there is a palpable tension 
between the observed mass of the Higgs boson and the stop mass. This is a version 
of the so-called ``little hierarchy problem'' \cite{Barbieri:2000gf}.  
Such an issue can be ameliorated 
in an extension of the MSSM called the next-to-MSSM (NMSSM) 
\cite{Ellwanger:2009dp} where the MSSM is 
augmented by an additional singlet scalar superfield. In such a scenario one can 
find, in contrast, tree level contributions to the Higgs masses that depend upon 
the new free parameters. Thus, the requirement of rather massive stops to obtain 
a $\sim 125$ GeV Higgs boson gets relaxed to a significant extent 
\cite{Chang:2005ht, Dermisek:2007ah, Kang:2012sy, Cao:2012fz, Agashe:2012zq}
and sub-TeV masses for both stops could easily serve the purpose. As far as 
the naturalness criterion is concerned, such a scenario conforms to the yardstick in 
a more efficient way than a scenario with only one relatively light stop 
\cite{Curtin:2014zua}. However, the issue of ``naturalness" is a more involved one 
and relates to other parameters in the theory in varied degrees. In the present 
study, we focus on such a scenario within the NMSSM where both stops are light 
(with sub-TeV masses).

On the other hand, the search for supersymmetric particles at the LHC has not 
only been quite elaborate but turns out to be quite exclusive as well in 
setting limits on the sparticle masses. The explicit constraints do however 
mostly restrict the strongly interacting sector which is expected due to their 
large production cross sections at a hadron collider like the LHC. Nonetheless, 
the third generation squarks present themselves with the weakest of the bounds 
from their direct searches. These happen mainly because of their smaller 
production cross sections when compared with the same for the squarks from 
the first two generations and also due to their possibly longer decay chains
which in turn weaken the signal strengths. Thus, rather light (sub-TeV) 
squarks from the third generation in a scenario like the NMSSM steals the 
spotlight and invigorates the program of dedicated studies for them at the LHC.

The paper is organized as follows. In section 2 we collect the basic
ingredients for our study where we present the situation with two light stops 
in the NMSSM in reference to the observed mass of the Higgs boson. We also 
discuss the neutralino sector of the NMSSM which is going to have a direct 
bearing in our present study. Section 3 is dedicated to finding the region of 
the NMSSM parameter space compatible with constraints originating from 
theoretical demands (like the ones arising by demanding absence of Landau poles 
in the evolutions of various new couplings), cosmological experiments shedding 
light on the dark matter (DM) sector, the experimental analyses in the Higgs 
sector and the direct SUSY searches at the LHC and other collider experiments. 
A thorough scan of the NMSSM parameter space is undertaken for the purpose and 
a few benchmark scenarios are chosen for simulation studies.  A detector-level 
simulation at the LHC is presented in section 4 for the benchmark scenarios 
along with a detailed simulation of the important SM backgrounds. We then 
discuss the prospects of discovering such light stops at the imminent LHC runs 
with higher center of mass energies. In section 5 we conclude.
%
\section{The situation with two light stops in the NMSSM}
In addition to the superfields appearing in the MSSM, NMSSM contains a singlet 
$\hat{S}$. In the following discussion we will confine ourselves to the $Z^3$ 
invariant NMSSM, i.e. we will ignore linear and bilinear terms in $\hat{S}$, 
as well as, the $\mu$ term. The superpotential reads \cite{Ellwanger:2009dp}
\beq
\mathcal{W}=\mathcal{W_{MSSM}}|_{\mu=0}+ \lambda \hat{S} \hat{H}_u.\hat{H}_d
  + {\kappa \over 3} \hat{S}^3,
\label{eq:suppot}
\eeq
where, $\cal{W_{MSSM}}$ is the superpotential for the MSSM, $\hat{H}_{u}$ and 
$\hat{H}_{d}$ denote the doublet Higgs superfields while $\hat{S}$ denotes the 
gauge singlet superfield. The corresponding soft supersymmetry-breaking terms 
are given by 
\beq
-\mathcal{L}_{\rm soft}= -\mathcal{L^{\rm MSSM}_{\rm soft}}|_{_{B\mu=0}}+ m_{S}^2 |S|^2 + 
\lambda A_{\lambda} S H_u.H_d  
+ \frac{1}{3} \kappa A_{\kappa} S^3 + {\rm h.c.} +.. \, 
\label{eq:ssbr}
\eeq
where, $\mathcal{L}^{\rm MSSM}_{\rm soft}$ denotes the soft supersymmetry-breaking 
terms in the MSSM; $A_{\lambda}$ and $A_{\kappa}$ are the soft supersymmetry-breaking 
terms with the dimension of mass; $m_S^2$ is the soft supersymmetry-breaking mass-squared term for 
$S$. During electroweak symmetry breaking (EWSB) the CP-even neutral components 
of $H_u$, $H_d$ and $S$, which are the scalar components of superfields 
$\hat{H}_u$, $\hat{H}_d$ and $\hat{S}$ respectively, assume vacuum expectation 
values ($vevs$) $v_u$, $v_d$ and $v_S$ respectively. Consequently, an effective 
$\mu$ term ($\mu_{\rm eff}$) is generated, which is given by 
$\mu_{\rm eff} = \lambda v_S$. This provides an elegant solution to the well-known 
``$\mu$-problem'' \cite{Kim:1983dt} that plagues the MSSM while the NMSSM was 
originally motivated over this virtue (see \cite{Ellwanger:2009dp} and references 
therein).
From our knowledge of the MSSM, where the Higgsino mass parameter $\mu$ plays a 
pivotal role in connecting the Higgs, the stop (and the bottom squark (sbottom)) 
and the electroweak gaugino/higgsino (electroweakino) sectors, it is not difficult 
to realize how NMSSM parameters could make their interplay even more involved.
Studies in the framework of the NMSSM discuss the situations with the Higgs 
sector
\cite{Stal:2011cz, Gunion:2012zd, King:2012is, Gunion:2012gc, King:2012tr,
Cao:2013gba, King:2014xwa}, 
the nature and role of the LSP
\cite{Ellwanger:1997jj, Ellwanger:1998vi, Dedes:2000jp, Choi:2004zx,
Cheung:2008rh, Stal:2011cz, Cerdeno:2013qta, Das:2012rr, Das:2013ta,
Kraml:2008zr, Panotopoulos:2011ew, Ellwanger:2013rsa, Ellwanger:2014hia}.
Among these, the more recent ones take into account the LHC-results on the Higgs 
boson and other relevant constraints.  Also, in the recent past, issues over the 
light stops and the sbottoms at the LHC \cite{Bi:2012jv, Guo:2013iij, Chakraborty:2015xia} 
in the context of the NMSSM and their implications for 
the 7 TeV (LHC-7) and 8 TeV (LHC-8) runs of the LHC (in addition to a relatively 
light gluino) have been discussed in some detail \cite{Cheng:2013fma}.

In the following subsections, we outline some relevant features of the three 
sectors mentioned above in the NMSSM framework before moving on to explore their 
implications in the context of the present work.

\subsection{The stop and the sbottom sectors of the NMSSM}
An NMSSM spectrum with two light stops inevitably contains at 
least one relatively light sbottom state. In particular, if the 
lightest stop ($\tilde{t}_1$) has a dominant left component, the 
presence of  a comparably light sbottom which is mostly left-chiral 
is automatic. This follows from the fact that the soft SUSY breaking mass 
term for the state $\tilde{b}_L$ has the same origin ($m_{\tilde{Q}_3}$) 
as that for the state $\tilde{t}_L$ and they differ only by the so-called 
$D$-term contributions which are relatively small and are model-independent 
for a given value of $\tan\beta$.
The mass-squared matrices involving the $\tilde{t}$ (in the basis $\tilde{t}_L$,
$\tilde{t}_R$) and $\tilde{b}$ (in the basis $\tilde{b}_L$, $\tilde{b}_R$) states 
in the NMSSM, (which, at tree-level, are similar to the ones in the case of MSSM) 
are given by \cite{Ellwanger:2009dp}
\beq\label{mstop} 
{\cal M}_{\tilde{t}}= \left(\ba{cc}  m_{\tilde{Q}_3}^2 + y_t^2 v_u^2 +
 (v_u^2-v_d^2)\left(\frac{g_1^2}{12}-\frac{g_2^2}{4}\right)
& y_t (A_t v_u - \mu_\mathrm{eff} v_d)  \\
 y_t (A_t v_u - \mu_\mathrm{eff} v_d) & 
m_{\tilde{U}_3}^2 + y_t^2 v_u^2-(v_u^2-v_d^2)\frac{g_1^2}{3}
\ea \right)
\eeq
and
\beq\label{msbottom}
{\cal M}_{\tilde{b}}=\left(\ba{cc} m_{\tilde{Q}_3}^2 + y_b^2 v_d^2 +
 (v_u^2-v_d^2)\left( \frac{g_1^2}{12}+\frac{g_2^2}{4}\right)
& y_b  (A_b v_d-\mu_\mathrm{eff} v_u) \\
y_b (A_b v_d-\mu_\mathrm{eff} v_u) & 
 m_{\tilde{D}_3}^2 + y_b^2 v_d^2 +(v_u^2-v_d^2)\frac{g_1^2}{6}
\ea \right)\; . 
\eeq  \\
Note that $\mu_{eff}(=\lambda v_S)$ replaces $\mu$ in the off-diagonal 
terms of these mass-squared matrices when compared to the MSSM case.
$m_{\tilde{Q}_3}$ stands for the soft SUSY breaking mass term for the 
third generation doublet (left-handed) squarks while $m_{\tilde{U}_3}$ and 
$m_{\tilde{D}_3}$ denote the same for the singlet (right-handed) stop and
sbottom states, respectively. $A_i$-s are the trilinear soft SUSY breaking 
terms and $y_i$-s denote the respective Yukawa couplings. $v_u$ and $v_d$ 
represent the $vevs$ of the CP-even up- and down-type neutral Higgs bosons,
$H_u^0$ and $H_d^0$, respectively and $g_2$ and $g_1$ denote the gauge 
couplings corresponding to $SU(2)_L$ and $U(1)_Y$ gauge groups, respectively. 
Smaller values of $m_{\tilde{Q}_3}$ and $m_{\tilde{U}_3}$ ensure lightness of 
both stops. Further, as has been explained earlier, small $m_{{\tilde Q}_3}$ 
corresponds to one light 
sbottom (which is dominantly left-handed). Proximity of $m_{\tilde{Q}_3}$ and 
$m_{\tilde{U}_3}$, and the largeness of $y_t$, 
guarantee significant mixing between the left- and the right-handed stops.
As a result, there may be a good amount of splitting between the two 
stop mass-eigenstates. This would then favor a hierarchy of low-lying stop and 
sbottom masses of the following kind: 
$ m_{\tilde{t}_1} < m_{\tilde{b}_1} < m_{\tilde{t}_2}$, 
which we study in some detail in the later part of this work\footnote{Hierarchies like
$\msbone \lesssim \mstone \lesssim \msttwo$ and $\mstone \lesssim \msttwo \lesssim \msbone$
are also possible depending upon the mutual hierarchy of $m_{\tilde{Q}_3}$ and
$m_{\tilde{U}_3}$.}. 
The chiral admixtures of the individual states depend on the hierarchy
of the soft masses $m_{\tilde{Q}_3}$, $m_{\tilde{U}_3}$ and $m_{\tilde{D}_3}$.

\subsection{The SM-like Higgs boson in the NMSSM}
\label{subsec:sm-higgs}
In the NMSSM, on EWSB, there are three CP-even Higgs 
bosons which are mixtures of CP-even (real) parts of $H_u$, $H_d$ and $S$. 
The LHC results now require one of these to be SM-like 
\cite{Flechl:2015foa}. 
The mass ($m_h$) of the SM-like Higgs boson ($h$) is given by 
\cite{Ellwanger:2011sk} 
\beq
\label{eq:hmass}
m_h^2 = m_Z^2 \cos^2 2\beta + \lambda^2 v^2 \sin^2 2\beta + \Delta_{\rm mix}+ 
\Delta_{\rm rad. corr.} \,
\eeq  
where $v = \sqrt{v_u^2 +v_d^2} \simeq 174 \, {\rm GeV}$, $\beta = \tan^{-1}
\frac{v_u}{v_d}$. 
The first term on the right hand side of this equation stands for 
the tree level squared mass of the Higgs boson 
in the MSSM. The second term is the NMSSM contribution at the tree level. 
The third term stems from singlet-doublet mixing which, in the limit of 
weak mixing, is given by
\beq
\label{eq:hmix}
\Delta_{\rm mix} = \frac{4 \lambda^2 v_S^2 v^2 
(\lambda -\kappa \sin 2\beta)^2} {\tilde{m}_h^2-m_{ss}^2}
\eeq
where $\tilde{m}_h^2 = m_h^2 - \Delta_{\rm mix}$ and 
$m_{ss}^2 = \kappa v_S(A_{\kappa}+4 \kappa v_S)$.  
As explored in reference \cite{Ellwanger:2011sk}, it is possible to raise 
the Higgs mass up to 140 GeV assuming perturbativity of $\lambda$ up to the GUT 
scale (i.e., no Landau pole is developed in the evolution of $\lambda$). If one
is ready to sacrifice such a requirement (the plausibility of which has recently
been discussed in the framework of a scenario like $\lambda$SUSY), it is possible
to have the doublet-like Higgs boson as heavy as $\sim 300$ GeV 
\cite{Ellwanger:2011sk,Barbieri:2006bg, Barbieri:2007tu}. 
Thus, even with a relatively small 
contribution from $\Delta_{\mathrm{rad. corr.}}$, the second and the third term could 
raise the tree-level Higgs mass significantly. However, note that $\Delta_{\rm mix}$
could have either sign and can increase or reduce the doublet-like Higgs mass.

On the contrary, in the MSSM, the lightest CP-even Higgs boson mass is bounded 
from above by $m_Z \cos 2\beta$ at tree-level. Hence, in such a scenario, 
a Higgs boson with mass $\sim 125$ GeV, as observed by the LHC experiments, 
requires a significant amount of radiative correction 
($\Delta_{\mathrm{rad. corr.}}$) which mostly arises at one-loop level with top 
quark and the stops in the loops thanks to the large top Yukawa coupling 
\cite{Okada:1990vk, Ellis:1990nz, Haber:1990aw}. For a fixed $\tan \beta$, 
with $m_A \gg m_Z$, the SM-like light CP-even Higgs mass [up to 
$\mathcal{O}\left(\frac{m_Z^2}{m_A^2}\right)$] is given by \cite{Djouadi:2005gj}
\beq
m_h  = \sqrt{(m_Z^2 \cos^2 2\beta + \epsilon \sin^2\beta)}
\left[1+ \frac{\epsilon m_Z^2 \cos^2\beta}{2 m_A^2 (m_Z^2 + \epsilon \sin^2\beta)} -
\frac{m_Z^2 \sin^2\beta +\epsilon \cos^2\beta}{2 m_A^2} \right]\
\eeq
where $m_A$ is the soft SUSY-breaking mass for the pseudoscalar Higgs boson and
the one-loop correction $\epsilon \simeq \Delta_{\rm rad. corr.}$ is given by
\cite{Carena:1995bx, Carena:1995wu, Haber:1996fp, Djouadi:2005gj}
\beq
\epsilon = \frac{3 m_t^4}{4 \pi^2 v^2 \sin^2 \beta} 
\left[2 \log \frac{M_S}{m_t} 
+ \frac{X_t^2}{ M_S^2} \left(1- \frac{X_t^2}{12 M_S^2} \right) \right].
\label{eq:radcorr}
\eeq
In the above expression, $m_t$ denotes the mass of the top quark\footnote{To take
into account the leading two-loop corrections to the Higgs mass, the running
$\overline{MS}$ top quark mass is to be taken \cite{Djouadi:2005gj}.}, 
$M_S=\sqrt{\mstone \msttwo}$ and $X_t=A_t-\mu\cot\beta$. 
It turns out that at least one of the $\tilde{t}$ states is required to be 
heavier than about a TeV and a large mixing between $\tilde{t}_L$ and 
$\tilde{t}_R$ is needed to push the (light) Higgs mass up to its 
experimentally observed value 
\cite{Arbey:2011ab, Draper:2011aa, Carena:2011aa, Carmi:2012yp, Arbey:2012dq, Bechtle:2012jw}.

As evident from the discussion above, in the NMSSM, in addition to the MSSM 
contribution, the Higgs mass receives extra contribution at the tree level 
as shown
in equation \ref{eq:hmass}. Thus, one does not necessarily bank on a heavy 
$\tilde{t}$ and/or large mixing to obtain a heavier Higgs boson
\cite{Ellwanger:2006rm, Barbieri:2006bg, Barbieri:2007tu}
with its mass around $\sim 125$ GeV 
\cite{Hall:2011aa, Arvanitaki:2011ck, Gunion:2012zd, King:2012is}
as observed by the LHC experiments.
This opens up, in the present context, the hitherto unexplored 
possibility of a theoretically consistent NMSSM scenario where both 
$\tilde{t}_1$ and $\tilde{t}_2$ have sub-TeV masses and can be quite 
light\footnote{Some such generic possibilities had earlier been discussed 
\cite{Curtin:2014zua} in the context of `natural' SUSY 
\cite{Papucci:2011wy} without trying to address the observed Higgs 
mass within 
the SUSY framework (thus implicitly relying on an NMSSM-like setup for
providing the extra contribution to the Higgs mass).}.
Such a provision warrants a 
thorough understanding of its implications, in particular, once it could
survive the current bounds on stop masses from the 
LHC experiments. In the following subsection we highlight the 
compatibility of having two relatively light stops in the NMSSM.
We also take a close look at its immediate implications for the scenario 
and contrast them with the situation in the MSSM.
%
\subsection{NMSSM versus MSSM: two light stops and the Higgs mass}
\label{subsec:neutralino}
%
To demonstrate how light the two stops could get to be, we scan over the 
NMSSM parameter space with the following ranges for various input parameters
using the popular package {\tt NMSSMTools} (v4.4.0) 
\cite{Ellwanger:2004xm, Ellwanger:2005dv, Ellwanger:2006rn, Das:2011dg, Muhlleitner:2003vg}:
%
\[
0.40 \leq \lambda \leq 0.75, \quad 0.01 \leq \kappa \leq 0.75, 
                      \quad |\mu_{eff}| \leq 1 \; \mathrm{TeV},  
                      \quad  1 \leq \tan\beta \leq 5, 
\]
\[
 |A_{\lambda}| \leq 2 \, \mathrm{TeV} ,  
\qquad
 |A_{\kappa}| \leq 2 \, \mathrm{TeV} , 
\qquad
|A_t| \leq \mathbf{3} \, \mathrm{TeV}, 
\]
%
\begin{equation}
 250 \, \mathrm{GeV} \leq (m_{_{\tilde{Q}_3}}, m_{_{\tilde{U}_3}}) \leq 3 \, \mathrm{TeV}.
\label{eqn:scan-range-nmssm1}
\end{equation}
%
The range of $\lambda$ ensures that we work in a large $\lambda$ regime for which 
the tree-level NMSSM contribution to the SM-like Higgs mass is appreciable. 
The range of Higgs mass that we allow in the scan is 
$125.1 \pm 3.0$ GeV\footnote{Note that the experimental uncertainty in determining 
$m_h$ is rather negligible ($\sim$ 240 MeV \cite{Aad:2015zhl}). 
The uncertainty of $\pm 3$ GeV takes into account the
theoretical uncertainty arising from missing higher order corrections in the
MSSM (see, for example, reference \cite{Carena:2013qia} and references therein). 
Additional corrections to the Higgs mass proportional to the parameter $\lambda$ 
at 2-loop have been calculated in reference \cite{Goodsell:2014pla}.  
The uncertainty interval can even be a little higher if one considers the 
uncertainty in the measurement of the top quark mass.}.
As can be seen from equation \ref{eq:hmass}, the tree-level NMSSM contribution 
to the Higgs mass is significant for small $\tan\beta$. Hence our choice for the
range of the same in equation \ref{eqn:scan-range-nmssm1}.

\begin{figure}[t]
\centering
\includegraphics[height=0.27\textheight, width=0.60\columnwidth , clip]{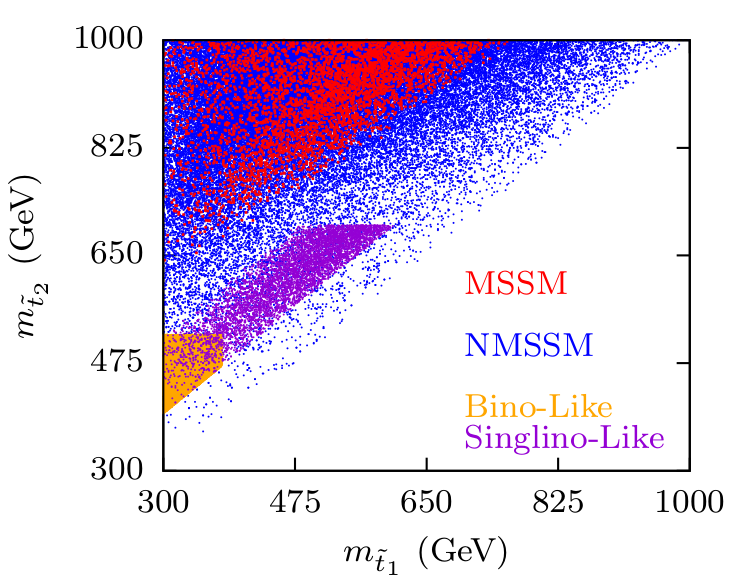}
\caption{Scatter plot showing the allowed regions in the $\mstone$-$\msttwo$ 
plane in the MSSM (in red; using {\tt FeynHiggs}) and in the NMSSM (in blue; 
using {\tt NMSSMTools}). Patches with yellow and purple points present
regions in the NMSSM parameter space under specific scenarios (bino- and 
singlino-dominated LSP neutralino, respectively) and satisfying some imposed spectral 
constraints discussed in section \ref{subsec:allowed}.
For the ranges of the model parameters scanned over 
in the respective scenarios, see text.}
\label{fig:compare-st1-st2}
\end{figure}

In figure \ref{fig:compare-st1-st2} we compare the NMSSM (in blue) and 
the MSSM (in red) in the $\mstone-\msttwo$ mass plane by scanning over the 
respective parameter spaces. For the MSSM, the results we present are by using 
{\tt FeynHiggs (v2.10.3)} 
\cite{Heinemeyer:1998yj, Heinemeyer:1998np, Degrassi:2002fi, Frank:2006yh, Hahn:2013ria}. 
However, we independently checked the same by running the popular spectrum 
generator {\tt SuSpect (v2.43)} \cite{Djouadi:2002ze}. We find close agreement 
between the two at the level of scanned output, except for some minor deviations 
arising out of known shifts in the SM-like Higgs mass by 2-3 GeV from the two
packages, for any given parameter-set. We also find very good agreements with
the existing literature \cite{Draper:2011aa, Carena:2011aa}. The ranges of 
various MSSM parameters employed in the scan are as follows:
%
\[
                 \quad |\mu| \leq 2.5 \; \mathrm{TeV},  
                 \quad  2 \leq \tan\beta \leq 32,   \nonumber
                 \quad |A_t| \leq \mathbf{3} \, \mathrm{TeV}, 
\]
%
\begin{equation}
 200 \, \mathrm{GeV} \leq (m_{_{\tilde{Q}_3}}, m_{_{\tilde{U}_3}}, m_A) \leq 2.2  \, \mathrm{TeV}.
\label{eqn:scan-range-mssm}
\end{equation}
For both the NMSSM and the MSSM scans, we fix $M_1=300$ GeV, 
$M_2=1.2$ TeV and $M_3=3$ TeV. The soft masses for all other squarks and 
sleptons are fixed at 3 TeV. We set the corresponding trilinear parameters 
$A_f$ to zero and the top quark mass $m_t$  to 173.1 GeV. The edge along
the diagonal traced out by the blue points marks the line below which
$\msttwo < \mstone$ which confronts their definitions.

It is clear from figure \ref{fig:compare-st1-st2}, for $\mstone$ as light as
300 GeV, the smallest $\msttwo$ that one can barely have in the MSSM is 
$\approx 650$ GeV. In contrast, in the NMSSM, $\sttwo$ can be as light as
$\approx 350$ GeV. In fact, the figure also reveals that a near-degenerate pair of 
stops is not impossible in the NMSSM, being still consistent with the 
constraint from the Higgs mass. Over the entire range of mass shown in the 
figure, for any given $\mstone$, NMSSM could provide us with a $\sttwo$ which 
can be lighter by $\approx 250-300$ GeV when compared to the MSSM.

\begin{figure}[t]
\centering
\includegraphics[height=0.22\textheight, width=0.49\columnwidth , clip]{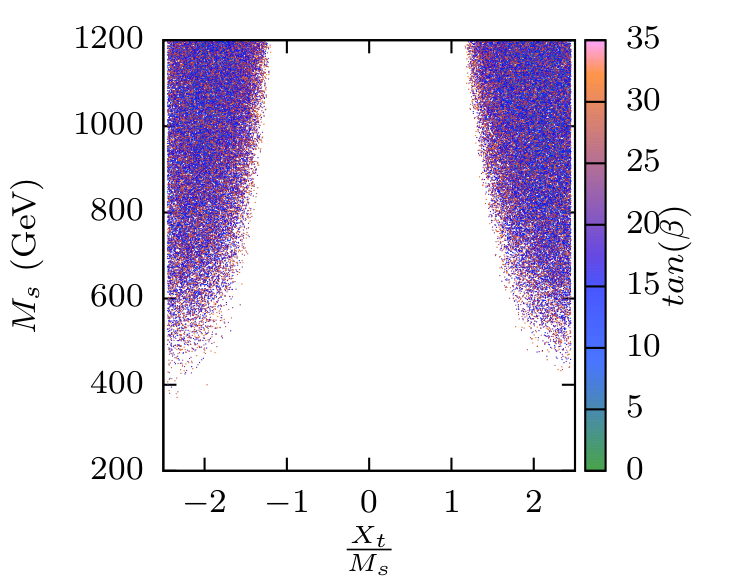}
\includegraphics[height=0.22\textheight, width=0.49\columnwidth , clip]{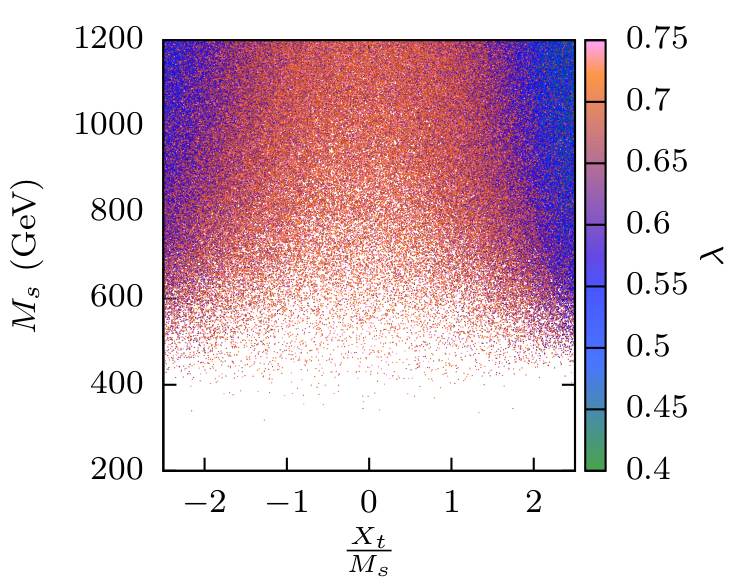}
\caption{Scatter plots showing the regions in the ${X_t \over M_S} - M_S$ 
plane compatible with the allowed range of the Higgs mass in the MSSM (left; using {\tt FeynHiggs}) 
and in the NMSSM (right; using {\tt NMSSMTools}). The color-code in the palette 
indicates the values of $\tan\beta$ and $\lambda$ for the left and the right plots,
respectively. The range of parameters scanned over in the respective scenarios are
same as in the case of figure \ref{fig:compare-st1-st2}. See text for details.}
\label{fig:compare-xt-ms}
\end{figure}

In figure \ref{fig:compare-xt-ms} we contrast the allowed regions in the 
${X_t \over M_S}$ plane for the MSSM (left) and the NMSSM (right). As shown, the 
color-code indicates the values of $\tan\beta$ (as is customary for the MSSM case)
and $\lambda$ in the respective
cases. It is 
evident from the left plot that in the MSSM, compatibility with observed mass of 
the Higgs boson requires either high values of $M_S$ or, in case of small $M_S$ 
(in turn, a small $\mstone$), rather large ${X_t \over M_S}$ (i.e., a large 
mixing parameter). Post Higgs discovery, this fact has become quite well-known 
and well-understood. 
Note that in the present work, we are interested in a pair of relatively light 
stops thus implying low $M_S$. For example, $\mstone \approx 300$ GeV with 
$\msttwo \approx 500$ GeV results in $M_S \approx 400$ GeV. It is clear from
figure \ref{fig:compare-xt-ms} that the MSSM could hardly afford such a spectrum,
and if at all, it would require a dangerously large value of ${X_t \over M_S}$
($\approx \pm \sqrt{6}$, that corresponds to the so-called ``maximal mixing 
scenario'', and is the range considered in figure \ref{fig:compare-xt-ms})
that could lead to a charge and color-breaking minimum 
\cite{Blinov:2013uda, Chowdhury:2013dka, Chattopadhyay:2014gfa, Camargo-Molina:2013sta, Camargo-Molina:2014pwa} for the scalar potential
of the theory or an unstable electroweak vacuum \cite{Blinov:2013fta, Bobrowski:2014dla}.
Furthermore, even for larger values of $M_S$, the required ${X_t \over M_S}$
remains moderately large in the MSSM resulting in two well-separated allowed
branches along the ${X_t \over M_S}$ axis.

In contrast, from the plot in the right, we find that in the NMSSM, the `desert' 
in the middle gets efficiently populated. This is since, in this region, 
Higgs mass gets 
significant tree level contribution in the NMSSM with $\lambda$ close to its 
largest value. Thus, unlike in the MSSM, a vanishing mixing parameter ($X_t$) in 
the NMSSM is very much compatible with the observed mass of the Higgs boson. 
However, in a scenario where both stops are relatively light and hence
the same for the soft masses for $\tilde{t}_L$ and $\tilde{t}_R$, it is not
automatic that a small $X_t$ would correspond to a small mixing angle $\theta_{\tilde t}$ 
in the stop sector. This is because in the limit of vanishing mixing parameter
the maximal mixing is still guaranteed if the diagonal terms of the ($2\times 2$) 
mass-squared matrix are equal. Given that we are interested in a scenario 
with two relatively light stops, the diagonal entries could be of comparable 
size. Thus, even with a vanishing $X_t$, a moderate mixing in the stop sector
is possible.

\begin{figure}[h]
\centering
\includegraphics[height=0.27\textheight, width=0.60\columnwidth , clip]{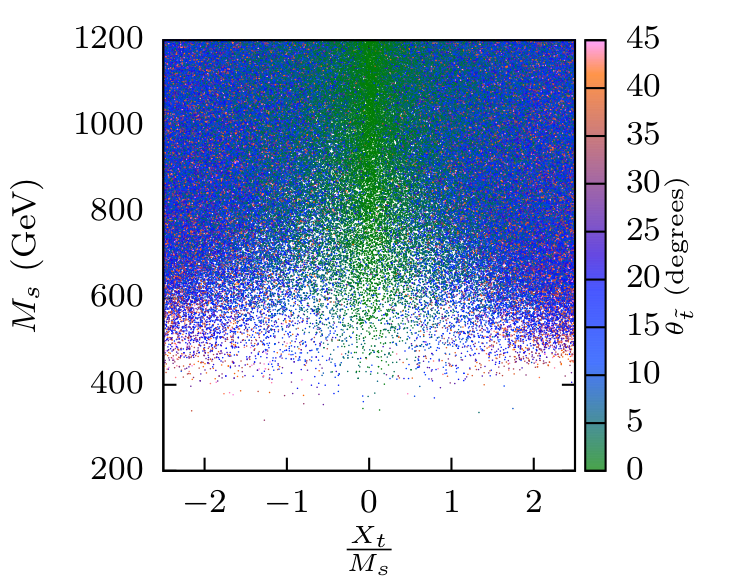}
\caption{Same as in the right plot of figure \ref{fig:compare-xt-ms} 
but with a palette indicating the mixing angle ($\theta_{\tilde{t}}$; in degrees) 
in the stop sector.}
\label{fig:mixing-nmssm}
\end{figure}

In figure \ref{fig:mixing-nmssm} we demonstrate this issue. Here we recast the right
plot of figure \ref{fig:compare-xt-ms} to indicate the amount of mixing, 
$\theta_{\tilde{t}}$, present in the stop sector.
We observe, that there is a preponderance of states with moderate to 
small mixings (in green and blue) for small $X_t$, with only a few isolated cases 
having relatively large mixing (in purple  and red). This is in sharp contrast 
to the MSSM scenario: in the NMSSM, it is possible to have a pair of relatively
light stops and that also with low mixing.
%

\subsection{The neutralino sector of the NMSSM}
\label{subsec:neutralino}
%
As in the case of MSSM, phenomenology of such light stop/sbottom squarks
of the NMSSM depends on the electroweak gaugino sector in
an essential way. In particular, the neutralino sector of the NMSSM 
differs from that of the MSSM in a crucial way. The fermionic component 
of the singlet superfield $\hat{S}$ in equation \ref{eq:suppot} 
(the `singlino', $\widetilde{S}$) could mix with the gauginos and the 
higgsinos of the MSSM. A singlino-dominated neutralino could turn out 
to be the lightest supersymmetric particle (LSP) and may crucially affect 
the cascade decays of the heavier SUSY excitations 
\cite{Ellwanger:1997jj, Ellwanger:1998vi, Dedes:2000jp, Choi:2004zx, Cheung:2008rh,
Stal:2011cz, Cerdeno:2013qta, Das:2012rr, Das:2013ta}. 
We would further assume that $R$-parity 
is conserved and hence the LSP ($\tilde{\chi}_1^0$) is stable.

The symmetric $5\times 5$ neutralino mass matrix (in the basis 
$\{\tilde{B},~\tilde{W}, ~\tilde{H}_1, ~\tilde{H}_2, ~\tilde{S}\}$)
is given by 
\beq\label{mneut}
{\cal M}_0 =
\left( \begin{array}{ccccc}
M_1 & 0 & -\dfrac{g_1 v_d}{\sqrt{2}} & \dfrac{g_1 v_u}{\sqrt{2}} & 0 \\
& M_2 & \dfrac{g_2 v_d}{\sqrt{2}} & -\dfrac{g_2 v_u}{\sqrt{2}} & 0 \\
& & 0 & -\mu_{\rm eff} & -\l v_u \\
& & & 0 & -\l v_d \\
& & & & 2 \kappa v_S
\end{array} \right)\; ,
\eeq
where $M_1$ and $M_2$ denote the soft SUSY-breaking masses for the 
$U(1)$ ($\widetilde{B}$) and the $SU(2)$ ($\widetilde{W}$) gauginos, respectively.
All other variables are described earlier in the text. Note that
the (1,5) and the (2,5) elements of the mass matrix are zeros. 
These imply that the singlino state does not mix directly with the 
$\widetilde{B}$ and the $\widetilde{W}$ states. Rather, its mixing 
to these states takes place indirectly via the higgsino sector.
It is thus natural to expect that gaugino-singlino mixings would 
never be too large.    
We consider two phenomenologically distinctive limiting possibilities with the 
neutralino LSP: (i) a bino-like LSP and  (ii) a singlino-like one. 
Such possibilities, along with the choice of a minimal (simplified) scenario 
appropriate for the purpose, lead us to some benchmark scenarios. These are 
discussed in the next section. We do not discuss the case of a higgsino-like 
LSP in this work. This is since, as elaborated in the next section, we like 
to keep the charginos out from the phenomenology we discuss
while a higgsino-like LSP would necessarily result in a light chargino. 
We postpone the discussion of such a possibility as well as 
the one that addresses a general situation of mixed electroweakinos 
to a future work.
%
\section{The compatible parameter space of the NMSSM}
%
The paradigm we are interested in is a `simplified' but an eminently
plausible scenario within the $Z_3$-symmetric NMSSM framework which is
consistent with the latest Higgs data and satisfies other important 
experimental constraints. Essentially, we have two relatively light stops, 
one light sbottom and a neutralino-LSP which can be either bino- or
singlino-like. We explore ranges where the lighter stop ($\stone$)
has mass below $\sim 550$ GeV while the heavier
stop is not heavier than, say, 700 GeV. The lighter sbottom can have
a mass in between the two stop masses. Such a spectrum of squarks
from the third generation have moderate to significant pair-production
cross sections at the imminent run of the LHC with increased center of 
mass energy and thus, is expected to be within its easy reach. Note that
(as mentioned in the previous section) the lighter chargino 
($\charonepm$) is taken to be rather heavy with $\mcharonepm > \msttwo$.

While exploring a spectrum with excitations as light as the ones 
mentioned above, it is imperative that one takes a critical 
note of relevant recent analyses reported by LHC-7 and LHC-8.
Indeed, all four 
light SUSY excitations indicated above draw crucial bounds from the 
latest data. We first discuss these bounds which lead us to a few 
benchmark scenarios appropriate for our present study.
%
\subsection{Experimental bounds on the relevant SUSY masses}
\label{subsec:expt-bounds}
Current experimental bounds from the LHC pertaining to the third 
generations squarks are grossly model-dependent in the
sense that they refer to specific mass hierarchies among the
involved states. 
It is thus important to check which of these bounds indeed apply for a 
scenario under study.

In the simplified scenario we consider, the heavier stop ($\sttwo$) is the
heaviest of the four relatively light SUSY states mentioned above.
We assume $\msttwo \sim m_t + \mntrlone$ such that the decay 
$\sttwo \to t \ntrlone$ is suppressed. Such a requirement is only relevant when 
the LSP is bino-dominated while for a singlino-like LSP such a decay is naturally disfavored. 
Thus, $\sttwo$ would have two
dominant decay modes: $\sttwo \to \stone Z$  and 
$\sttwo \to \sbone \wplus/{\wplus}^*$.\footnote{A stringent requirement
of $\mstone + m_Z \leq \msttwo \leq m_{\mathrm{top}} + \mntrlone$ would imply
$\mstone - \mntrlone < m_{\mathrm{top}} -m_Z (\approx 82$ GeV) which is smaller
than $m_b + m_W$. This would ensure $\stone$ always have a 
flavor violating (FV) decay to $c \ntrlone$ and/or four-body (4B) decay
to $b f \bar{f}^\prime \ntrlone$.}
It is also possible that $\sttwo$ decays to $\stone$ and a light
Higgs boson when kinematically allowed.
By restricting the mass-split between $\sttwo$ and $\ntrlone$ in the
above fashion, we end up with somewhat compressed spectra comprising 
of closely spaced states like $\sbone$, $\stone$ and the LSP below the 
state $\sttwo$.
\begin{figure}[htb]
\centering
\includegraphics[height=70mm, width=\textwidth]{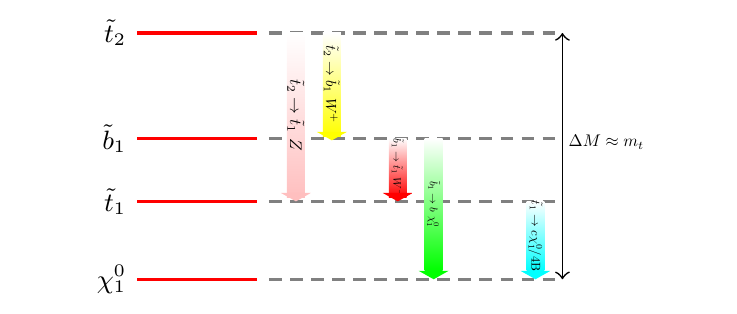}
\vskip -10pt
\caption{The mass-hierarchy of the lowest lying excitations that include two
light stop and a light sbottom. The LSP $\ntrlone$ can either
be a bino or a singlino. The possible decay modes of these states
are also shown. The mass-split between $\sttwo$ and $\ntrlone$ 
($\Delta M$) is 
required to be around the mass of the top quark for the simplified scenario
discussed in this work. Note that the relative splittings shown in the figure
are only representative and are not to scale.}
\label{fig:cascade}
\end{figure}

As for $\sbone$, in general, the two-body decay modes $\sbone \to b \ntrlone$ 
and $\sbone \to \stone \wminus$ may both be substantial if kinematically 
allowed. However, if the LSP is singlino-dominated, the coupling 
$b \sbone \ntrlone$ will be suppressed and the decay $
\sbone \to \stone \wminus$ would prevail. Nonetheless, given that the 
mass-split between the states $\sbone$ and $\stone$ may be naturally 
small in some regions of the SUSY parameter space, the latter mode
(via on-shell or off-shell $W$-boson, as the case may be) might 
also  experience a phase-space suppression resulting in a competition 
between these two available modes, as long as the singlino admixture in
the LSP is not too large.

The lighter stop, in our scenario, could only undergo the loop-level  
flavor-violating
decay $\stone \to c \, \ntrlone$ \cite{Hikasa:1987db}\footnote{A more recent 
work performs the complete one-loop calculation of the decay 
$\stone \to c \ntrlone$ \cite{Muhlleitner:2011ww} in the
framework of minimal
flavor violation (MFV). It is reported that a deviation of about 10\% is 
typical for large values of the MFV scale while the same could be bigger if
the MFV scale is small. We use {\tt NMSSMTools} for our study which only has the
results of reference \cite{Hikasa:1987db} implemented.} and/or a decay to a 
four-body final state $\stone \to b  f \bar{f}^\prime \ntrlone$ 
\cite{Boehm:1999tr, Das:2001kd} (as $\mstone - \mntrlone < m_b + m_W$). 
Thus, bounds from only those experimental analyses that considered these 
possibilities would be relevant for the present study. We now briefly discuss 
the current experimental situation below.

Interestingly, direct production of a pair of $\sttwo$ and their decays to $\stone Z$ 
have been discussed for the first time by both the ATLAS and the 
CMS collaborations \cite{Aad:2014mha, Khachatryan:2014doa} only very recently. 
The latter,
in addition, considers the decay mode $\sttwo \to \stone h$. A phenomenological analysis 
of such a decay has been performed in reference \cite{Ghosh:2013qga}. It is to be noted 
that the decay 
mode $\sttwo \to \sbone \wplus$ which is relevant for the scenario we are 
discussing in this work, has not yet been considered in the experimental 
studies to the best of our knowledge. In any case, the analyses mentioned 
in these works
assume that $\mathrm{BR}\left[\stone \to t \ntrlone \right]$ is 100\% which is
kinematically disfavored
in our scenario. Thus, the bounds obtained there are not applicable to 
our present study. Constraint on $\sbone$-$\ntrlone$
mass-plane has been derived in reference \cite{Aad:2013ija, cms-sbot-1} 
assuming 
$\sbone$ always decaying to $b \ntrlone$. As noted earlier, for varied 
reasons, this may not be the most favored mode of decay for $\sbone$ 
in our case and hence the constraint would be relaxed. Reference \cite{Aad:2014nra} 
presents dedicated studies on the production of a pair of $\stone$ 
followed by their FCNC-decays to charm quarks and the LSP using both
charm-tag and monojet-like selection (a hard jet originating in the 
initial state radiation). The latter technique is also employed in
reference \cite{cms-stop-1} to constrain the $\mstone$-$\mntrlone$ plane
for small mass-split between these two states. 
Again, in references \cite{Aad:2014nra, Aad:2014kra}, such a monojet-like search
is adopted to probe the parameter space with highly mass-degenerate 
$\stone$ and $\ntrlone$ that leads to four-body decays of $\stone$
discussed earlier. Table \ref{tab:refs} serves as a ready-reckoner for these
experimental analyses.
\begin{table}[!htbp]
\centering
\begin{tabular}{ |c|c|c| }
\hline \hline
   Decay modes & ATLAS & CMS \\
\hline \hline
$\sttwo \to \stone Z$ & \cite{Aad:2014mha} & \cite{Khachatryan:2014doa} \\
$\sttwo \to \stone h$ &  Not available     & \cite{Khachatryan:2014doa} \\
$\sttwo \to \sbone W^+$ & Not available    & Not available \\
\hline
$\sbone \to b \ntrlone$ & \cite{Aad:2013ija} & \cite{cms-sbot-1} \\
$\sbone \to \stone \wminus$ & Not available & Not available \\
$\stone \to c \ntrlone$ & \cite{Aad:2014nra} & \cite{cms-stop-1} \\
\hline
$\stone \to b f \bar{f}^\prime \ntrlone$ & \cite{Aad:2014kra, Aad:2014nra} & 
Not available \\
\hline
\end{tabular}
\caption{List of references for various experimental analyses at the LHC
relevant for the present work. See text for the
constraints extracted from these references.
}
\label{tab:refs}
\end{table} 

\subsubsection{Scenario with a bino-like LSP}
In the NMSSM, the lightest neutralino (LSP) is bino-like when the 
soft-breaking $U(1)$ gaugino mass parameter ($M_1$) is much smaller than 
both $\mu_{eff}$ and the NMSSM quantity $\kappa v_S$. 
For such a neutralino the bounds derived in reference 
\cite{Aad:2013ija, cms-sbot-1}
will be applicable. Thus, the smallest LSP mass that we could legitimately
consider is dictated by the mass of the lighter sbottom ($\msbone$) 
we would use. A ballpark value of $\msbone \approx 400$ GeV would require 
$\mntrlone \gtrsim 270$ GeV. Once compatible ranges for $\msbone$ and 
$\mntrlone$ are chosen, we construct the benchmark scenarios by requiring 
the proposed hierarchy of masses: 
$\mntrlone < \mstone < \msbone < \msttwo < \mcharonepm$.
The bounds $\mstone > 240$ GeV for arbitrary $\mntrlone$,
$\mstone > 270$ GeV for $\mntrlone=200$ GeV and the one for the nearly
degenerate case $\mstone \approx \mntrlone > 260$ GeV \cite{Aad:2014nra} 
all hold in such a scenario.
%
\subsubsection{Scenario with a singlino-like LSP}
%
A singlino-like neutralino-LSP is realized in the NMSSM for small values of 
$\kappa v_S$ \cite{ Das:2012rr, Kraml:2008zr, Ellwanger:2013rsa, Ellwanger:2014hia}.
The couplings of such an LSP state to any other excitation is generically 
suppressed. Thus, for such an LSP, the decay rate in the mode 
$\sbone \to b \ntrlone$ can be significantly suppressed. This may lead to 
either a long-lived/metastable $\sbone$ (if this is the only decay mode which 
is kinematically allowed) or to a reduced branching fraction for the same 
(in the presence of its other available decay modes). In any case, under such 
a circumstance, the collider-bound on the mass of the LSP that was relevant for 
the bino-like case (a prompt $\sbone \to b \ntrlone$ decay with 100\% branching 
fraction) is not applicable. 
In sharp contrast to the bino-dominated LSP case, in our present scenario, 
$\sbone$ would dominantly decay via $\stone W/W^{*}$. Unfortunately, as 
mentioned earlier, no experimental analysis seems to exist that exploits this 
particular decay mode to put bounds on the sbottom 
mass\footnote{Phenomenology of such a decay mode at the LHC has been discussed
in references \cite{Li:2010zv, Datta:2011ef}}. Under the 
circumstances, to be conservative, we continue to consider $\mntrlone > 270$ 
GeV. 
For a singlino-like LSP, both flavor-violating two-body decay 
$\stone \to c \ntrlone$ and four-body decay of $\stone$ would get further 
suppressed. However, as long as the total (combined) decay width 
($\Gamma_{\mathrm{tot}}^{(\stone)} \gtrsim 10^{-13}$ GeV) ensures an effectively prompt 
decay of $\stone$, the bounds on 
$\mstone$ based on such a decay \cite{Aad:2014nra}, as described above in the 
case of a bino-dominated LSP, apply. We would briefly look 
into this kind of a critical situation and its possible implications in section
\ref{subsec:cascade-stop2}.
%
\subsection{Allowed regions of the NMSSM parameter space}
\label{subsec:allowed}
%
In order to have an idea how big the NMSSM parameter space consistent
with the simplified scenario proposed in figure \ref{fig:cascade} and
compatible with some theoretical and current experimental 
constraints is, we undertake a thorough scan of the same using
{\tt NMSSMTools}. 
The ranges for various parameters scanned over are taken from equation
\ref{eqn:scan-range-nmssm1}.
The Fortran code {\tt NMHDECAY} 
\cite{Ellwanger:2004xm, Ellwanger:2005dv, Belanger:2005kh} 
in {\tt NMSSMTools} computes the masses, couplings and decay widths 
of all Higgs bosons of the NMSSM, and the masses of all sparticles.
Again, the range of Higgs mass allowed for the scan is $m_h=125.1 \pm 3.0$ GeV, as is 
default to {\tt NMSSMTools}.
As discussed in the previous subsection, we ensure the mass of the LSP neutralino 
to be around its lowest acceptable value compatible with the overall scenario we 
adhere to, i.e., $\sim 300$ GeV. For the bino-dominated LSP, we thus fix 
$M_1=300$ GeV. In the case of singlino-dominated LSP, such a mass for the LSP is 
obtained for relatively smaller values of $\kappa$ as the latter varies. Thus, 
for this case, we fix $M_1$ at 700 GeV. Further, to have a lighter chargino 
heavier than the two stop states and the lighter sbottom, 
we take $M_2$ large enough ($\sim 1.2$ TeV). On the same ground,  only a 
large enough value of $\mu$\footnote{Note that by such a 
choice we adopt a scenario where $M_1 << \mu \, , M_2$ thus making way for either 
a bino-dominated or a singlino-dominated neutralino LSP depending on the value 
of the parameter $\kappa$.} would be compatible with our proposed scenario. 
Also, $M_3$, the masses of the sleptons and those of the squarks of the first 
two generations are held fixed at 3 TeV which help evade the LHC bounds on the 
masses of the gluino and the sfermions straight away. As far as benifitting quantitatively 
in terms of the degree of finetuning, we find that, within the framework of NMSSMTools \cite{Ellwanger:2011mu,Ellwanger:2006rn}, 
the finetuning parameter ($\Delta$) is in the ballpark of $\simeq 10 (30)$ for
$\mstone=300 \, (500)$ GeV and $\msttwo \approx \mu=450 \, (650)$ GeV, which are
compatible with the simplified scenario we consider in this work. The values of $\Delta$ 
could be compared with the typical MSSM values which are at least a few hundred.
If one is ready
to give up on the simplified scenario, a smaller $\mu$ is acceptable thus
improving further on the degree of finetuning.

%

To adhere to the simplified 
scenario proposed in figure \ref{fig:cascade}, we require that $\sttwo$ does not 
have any appreciable branching fraction to $t \ntrlone$. Thus, during the present
scan, we ensure $\msttwo - (m_t + \mntrlone) < 50$ GeV. 
Furthermore, as mentioned in section \ref{subsec:expt-bounds}, we ensure
that $\msttwo - \mstone > m_Z$ so that $\sttwo$ could decay to $\stone Z$. Also,
we restricted $\mstone - \mntrlone$ to $<80$ GeV so that the only decays of 
$\stone$ are to flavor-violating two-body mode $c \ntrlone$ and/or four-body
$b f \bar{f}^\prime \ntrlone$ mode.
The scan is subjected to various phenomenological constraints that are 
in-built in {\tt NMSSMTools}. To name a few (see the webpage mentioned
in reference \cite{Ellwanger:2004xm}), these are the constraints from
$B$-physics, radiative Upsilon decays, ALEPH constraints on 
$H \to AA \to 4\tau$, Tevatron and LHC constraints on the charged Higgs 
sector, LHC constraints on $H/A \to \tau^+ \tau^-, Z \gamma$, 
$H \to bb,ZZ, WW, \gamma\gamma$, $H \to AA \to 4 \mu$ and the 
LHC bound on the Higgs mass.

At this point a little digression to figure \ref{fig:compare-st1-st2} 
would help understand where exactly our scenario lives in an otherwise allowed region
of the NMSSM parameter space. The combined band in yellow (bino-like LSP neutralino) 
and purple (singlino-like LSP neutralino) is the relevant region. The lower (upper) 
diagonal edge arises by demanding $\msttwo \geq \mstone + m_Z$ 
($\msttwo - (m_t +\mntrlone)  \leq 50$ GeV). For the bino-like LSP case, the LSP mass
 is more or less fixed at $\approx 290$ GeV, being determined by the choice $M_1=300$ GeV.
Consequently, $\msttwo$ is bound from above and this is reflected in the flat edge at the
top of the yellow band. On the other hand, for the singlino-like case, the LSP mass
is determined by $\kappa v_S$ and hence it varies in our scan. This pushes up the values
of $\mstone$ and $\msttwo$ that are consistent in our scenario. The flat edge at the 
top of the purple band relates to the maximum value of $\msttwo$ ($\leq 700$ GeV) 
that we allow. Clearly, this is an artificial cut-off but a meaningful one that 
restricts us only to a relatively light pair of stops.

A SUSY explanation of the tantalizing excess (at $\sim 3.5 \sigma$ level)
in the measured value of muon $(g-2)$ \cite{Bennett:2006fi, Blum:2013xva} over its SM 
prediction calls for a small smuon mass and a somewhat large $\tan\beta$,
in particular, if the charginos are heavy. 
On the other hand, in our scenario with large 
values of $\lambda$ (and light stops), only small $\tan\beta$ 
($1 \lesssim \tan\beta \lesssim 5$) 
complies with the allowed range of Higgs mass. Hence it is difficult
to satisfy the 
muon $g-2$ constraint \cite{Endo:2013bba, Badziak:2014kea} from within 
{\tt NMSSMTools} (which allows for only a $2\sigma$ window about the measured central 
value) without requiring the smuon so light that it becomes the LSP.
However, allowing for a $\sim 2.5 \sigma$ 
downward fluctuation (i.e., only a smaller
excess over the SM) could easily accommodate a 
suitably light smuon. Note that a light smuon does not enter the 
phenomenology we 
discuss in this work unless a chargino or a heavier neutralino is
lighter than, at least, the heavier stop. Furthermore, such a scenario
but with a heavier spectrum 
would allow for a larger $\tan\beta$. In that case a heavier smuon
could fit in. For simplicity,  we fix the smuon mass to a high value in this work.

In the DM sector, two different experimental constraints are in reference: 
(i) the Planck result that restricts the range of allowed relic abundance 
($\Omega_c h^2 \simeq 0.119 \pm 0.002$) \cite{Ade:2015xua} and 
(ii) the constraint on the maximum allowed spin-independent DM-nucleon 
elastic scattering cross section (as a function of the mass of the DM 
candidate) as reported by the direct detection experiment LUX 
\cite{Akerib:2013tjd}. These constraints can be evaded by going beyond the 
standard cosmological model and/or assuming non-thermal 
DM.\footnote{For 
example, in the presence of non-thermal production of the DM candidate, 
the constraint on the DM annihilation cross-section obtained by assuming 
thermal production of DM can be evaded 
\cite{Moroi:1999zb, Acharya:2008bk, Acharya:2009zt}. 
Further, the presence of a lighter gravitino or axino would invalidate the 
constraints on the lightest neutralino as the DM candidate. In the presence 
of other DM components (from a hidden sector, for example), the same 
constraints would again be relaxed. All these possibilities would make 
little or even no difference to the collider signatures we discuss in this 
work (and may not even have their own distinctive signatures).}
Nonetheless, a somewhat relaxed version of the constraint on the relic 
abundance may be to respect only the upper bound on the same thus ensuring 
that the universe is not over-closed. We also observe that the LUX
bound is easily satisfied almost over the entire region of the parameter 
space of our interest, for both bino- and singlino-dominated LSP cases. Rather
heavy ($\sim 3$ TeV) squarks from the lighter generations help keep the 
spin-independent DM-nucleon elastic scattering cross section low enough for the
purpose.
As for the constraint related to the non-appearance of Landau pole (up to the grand unification scale $\sim {\cal O}( 10^{16})$ GeV ), this again can be overlooked in the scheme of the so-called $\lambda$SUSY 
\cite{Barbieri:2006bg, Hall:2011aa, Hardy:2012ef, Farina:2013fsa} where large 
values of $\lambda$ (up to 2) at the weak scale can be allowed. 

\begin{figure}[htb!]
\centering
\includegraphics[height=0.75\textheight, width=\textwidth]{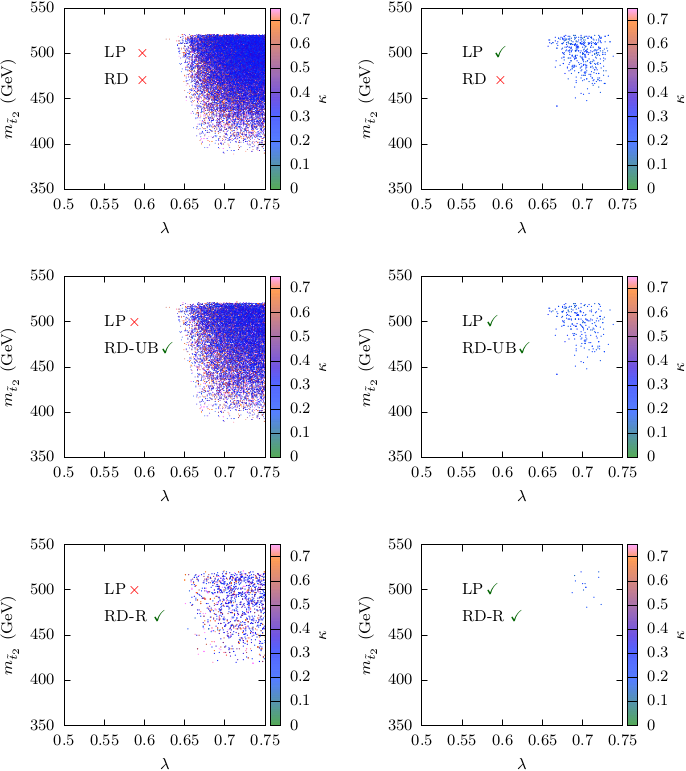}
\caption{Scatter plots depicting the regions in the $\lambda$-$\msttwo$ plane 
compatible with the simplified scenario we consider and for the case of a
bino-dominated LSP. All major electroweak constraints (including those for 
the Higgs-sector observables) as incorporated in the package NMSSMTools are 
considered except for the one related to muon $(g-2)$. The values of 
$\lambda$ and $\kappa$ are at the fixed scale 
of the common soft mass of the light-flavored squarks. The legends `RD', 
`RD-UB' and `RD-R' stand for `relic density', `relic density upper bound' 
and `relic density range', respectively. The legend `LP' stands for 
`Landau pole' related requirements. The symbols `$\checkmark$' and `$\times$' 
against these legends convey whether these particular types of constraints are 
imposed or not. See text for details.}
\label{fig:rdlp-bino}
\end{figure}

\begin{figure}[htb]
\centering
\includegraphics[height=0.75\textheight, width=\textwidth]{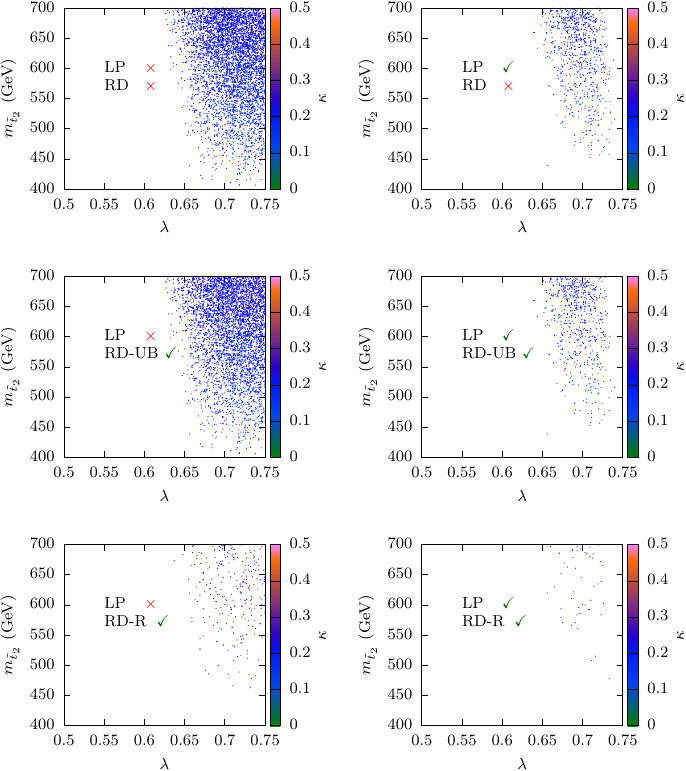}
\caption{Same as in figure \ref{fig:rdlp-bino} but for the case of a
singlino-dominated LSP.}
\label{fig:rdlp-singlino}
\end{figure}

In the light of the above discussion, we adopt the following strategy to 
understand the impacts of incorporating the constraints pertaining to
relic-abundance and Landau pole only. 
In figure \ref{fig:rdlp-bino} and \ref{fig:rdlp-singlino} we present the 
results of some kind of a ``constraint-flow'' analysis for the cases with 
bino- and 
singlino-dominated LSP, respectively. These demonstrate the outcomes of 
imposing these constraints successively thus shedding light on their
respective impacts. In the case of the allowed range for relic abundance
we use the one incorporated in the NMSSMTools, i.e., 
$0.107 \leq \Omega_c h^2 \leq 0.131$. The plots in these two figures are drawn in the 
$\lambda$-$\msttwo$ plane with the values of the parameter $\kappa$
represented by the color-code defined in the adjacent color-palettes. 
In both figures, the plots on the top, left corners delineate the 
regions allowed after imposing all the constraints (including the
experimentally allowed range of the mass of the Higgs boson) discussed 
earlier except for the ones related to the DM relic abundance 
and the Landau pole. This plot in each figure serves as the `reference' 
with respect to which the effects of the latter two constraints are 
studied.

For the bino-dominated LSP (figure \ref{fig:rdlp-bino}) we require the 
bino-admixture in the LSP to be $\geq 95\%$. The reference plot clearly 
reveals that rather light stops are 
only compatible with large values of $\lambda$. This is just an artifact 
of the basic paradigm we are exploiting in this work, i.e., larger 
$\lambda$ compensates for lower stop masses to have the 
mass of the SM-like Higgs boson within the experimentally allowed range.
Meanwhile, $\kappa$ can attain values ranging over moderately low to the 
highest value allowed in our scan. The sharp, flat edge at the top of 
the region indicates the largest mass for $\sttwo$ which is consistent 
with the scenario we are working in 
(i.e., $\msttwo - \mntrlone \leq m_t + 50 \; \mathrm{GeV}$).

Similarly, for a singlino-dominated LSP (figure \ref{fig:rdlp-singlino}) 
we demand the singlino-component in the LSP should be $\geq 95\%$. The 
reference plot has more or less a uniform density for 
$\lambda \gtrsim 0.65$ and 
$400 \, \mathrm{GeV} \lesssim \msttwo \lesssim 700 \, \mathrm{GeV}$.
Low values of $\kappa$ (up to $\kappa \simeq 0.3$) are only consistent
since these naturally render the LSP singlino-dominated as long
as $\kappa v_S=\kappa \mu_{eff}/\lambda << M_{1,2}$.
Note that as $\kappa v_S$ increases within this range, the mass of the
singlino dominated LSP also increases. This is in contrast with the
bino-dominated LSP case described earlier where the LSP mass gets more
or less fixed by fixing the MSSM parameter $M_1$. An increasing
mass of the LSP drives (slides) the entire spectrum of our simplified scenario
to the heavier side. 

We now move on to impose the constraint from the DM relic abundance as reported 
by the Planck experiment. In addition, we would also demand absence of Landau
poles in the evolution of the parameters like $\lambda$, $\kappa$ and the top 
and the bottom quark Yukawa couplings up to the unification scale which is around $10^{16}$ GeV. 
Along the rows we show the allowed regions on further impositions of the relic density 
upper bound of $\Omega_ch^2=0.131$ (the loose criterion; second row) 
and the constrained range for the same as implemented in 
{\tt NMSSMTools} and as mentioned earlier (the tight 
criterion; last row). On the other hand, for the plots in the first column, the 
Landau pole-related constraint is not imposed while for the ones in the second 
column this is incorporated. 

It can be clearly seen that as we go from the top to the
bottom of these figures, the `loose' demand on the relic abundance does 
not affect the otherwise allowed regions of the NMSSM parameter plane in 
any significant way. However, incorporating the `tight' criterion on the
relic abundance results in shrinking of the allowed parameter plane 
towards larger values of $\lambda$ and $\msttwo$ in the case of a 
bino-dominated LSP. That the smaller values of $\msttwo$ get 
less preferred is understandable as follows. Allowing for a range of relic abundance,
which is now also bounded from below, implies that we cut off the low-abundance 
region, i.e., the region with
higher overall annihilation rate. This is achieved when the stop masses are increased 
since these reduce the LSP-$\tilde{t}_1$ coannihilation rate in the one hand and the 
$\tilde{t}_1$ and $\tilde{t}_2$ mediated LSP annihilation on the other.

As for the singlino-dominated LSP, a careful study reveals 
that the `tight' criterion does not alter the already-allowed region too
much. The only effect it has is in the form of lowering the
density of points in the allowed region (which is a natural effect when 
a new constraint is imposed). This may be understood in terms of the fact that 
unlike in the bino-like LSP case, dominant annihilation and coannihilation modes 
of the singlino-LSP do not involve $\tilde{t}_2$. Hence, as the `tight'
criterion does, putting a lower bound on the relic abundance does not have 
much effect on the already allowed region of the NMSSM parameter plane.

Next we discuss the impact of imposing the demand related to Landau pole 
(as we move from the first column to the second). In both bino- and 
singlino-dominated LSP cases, this turns out to be more restrictive 
compared to the ones pertaining to the relic abundance. Irrespective of 
the nature of the LSP, this constraint puts an upper bound on 
$\lambda$.\footnote{The maximum value of $\lambda$ that we allow in our 
scan ($\lambda=0.75$, at the SUSY breaking scale given by 
the average mass of the squarks from the first two generations, i.e.,
3 TeV) is, in the first place, motivated by the 
Landau-pole constraint.  As discussed earlier in this section, a scenario 
like $\lambda$SUSY considers large values of $\lambda$ (up to 2) at the 
weak scale without jeopardizing the unification of gauge couplings at a 
high scale \cite{Hardy:2012ef}.} 
For a bino-dominated LSP, the Landau pole constraint pushes 
up the stop masses by an appreciable extent.

\begin{figure}[htb]
\centering
\includegraphics[width=\textwidth]{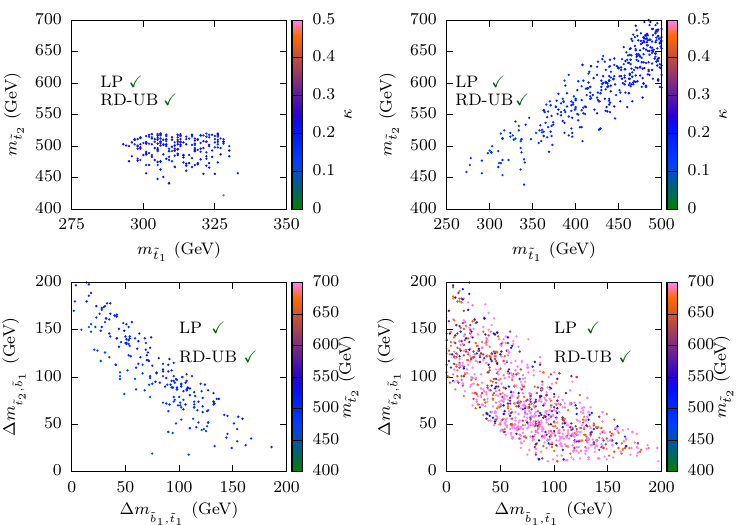}
\caption{Scatter plots depicting the regions in the $\mstone$-$\msttwo$ plane 
(upper panel) and in the plane of two mutual splittings: 
$\Delta m_{_{\sbone,\stone}}$ and $\Delta m_{_{\sttwo,\sbone}}$ (lower panel). 
As before, the left column represents the case of
bino-dominated LSP while the right column does the same for singlino-dominated
LSP case. In each plot and for each scatter point, the value of $\kappa$ 
(upper panel) or $\msttwo$ (lower panel) is indicated by the color code defined in
the adjacent color-palette.
For all the cases, the scans are subjected to the constraint pertaining to the 
absence of Landau pole and that on the maximum of DM relic density.
}
\label{fig:rdlp-others}
\end{figure}

In figure \ref{fig:rdlp-others} we present the projected regions in different 
parameter planes which are relevant for the current study. We opt for
a reasonably conservative scenario where the Landau pole related 
constraint is enforced in the strictest sense but only the upper bound, 
rather than the allowed range, of the relic abundance is required to be 
satisfied. Plots in the left column present the cases with bino-dominated 
LSP while the ones in the right column illustrate the cases where the LSP 
is singlino-dominated. 
The top row delineates the compatible region in the $\mstone$-$\msttwo$ in
our simplified scenario, with variation of $\kappa$ being indicated by 
the color-code defined in the adjacent palettes. The plot on the top, 
left corner have the same set of points as the plot in the second row
and second of column of figure \ref{fig:rdlp-bino} but now projected
on a different plane. The points here populate a roughly rectangular box
whose edges are determined by the hierarchy we have adopted for our 
scenario, the mutual mass-splits that we have allowed and the minimum 
LSP mass that we have chosen. On the other hand, the right plot in top
row presenting the case with singlino-dominated LSP, clearly displays
a correlation between $\mstone$ and $\msttwo$ as they grow. It has been
already discussed that allowing for a variation of $\kappa$ implies a
change in mass of the LSP-singlino which `slides' the entire spectrum
to high mass-values. The almost uniform width of the band along the
diagonal results from a very restrictive mutual splittings that are
enforced among the low-lying states to comply with our simplified
scenario. One may like to refer back to the yellow (bino-dominated LSP)
and purple (singlino-dominated LSP) bands in figure \ref{fig:compare-st1-st2} 
essentially from which
the left and the right plots in the first row 
of figure \ref{fig:rdlp-others} are derived by
imposing further constraints. 
Also, note that in both the cases $\kappa$ remains small. 
In the bino-dominated (left) case, $\kappa$ could have 
been larger were it not for the fact that it is prohibited by
demanding absence of Landau pole.

From the plots in the bottom
row we clearly see a strong correlation between the mutual mass-splits:
an increasing split between $\msbone$ and $\mstone$ results in a shrinking 
of the split between $\msttwo$ and $\msbone$. This is natural because of
the imposed hierarchy $\mstone < \msbone < \msttwo$ with a restricted
maximum splitting between $\msttwo$ and LSP mass. Since the LSP mass is
essentially determined by the fixed value of  $M_1$ ($=300$ GeV) in the 
bino-dominated case, the left figure displays a uniform (blue) value
that $\msttwo$ could take as all other parameters vary in the scan.
On the other hand, for the reason discussed above, the case with 
a singlino-dominated LSP allows for a range of $\msttwo$ (blue and
reddish points) as $\kappa$ varies. One can notice that somewhat lower
mutual splittings are allowed simultaneously in this case.

%
\subsection{Choosing the benchmark scenarios}
\label{subsec:benchmarks}
%
In table \ref{tab:benchmarks}, we present three benchmark scenarios that
satisfy the simplified situation demonstrated in figure \ref{fig:cascade}: 
two of them are for the bino-like LSP case and one is for the singlino-like 
LSP case. 
All three scenarios are confronted by the constraints coming
from various different search modes at the LHC experiments via the package 
{\tt CheckMATE (v1.2.0)} \cite{Drees:2013wra} (to the extent the corresponding analyses
are available with the said package) and are found to pass them.
These points also satisfy the upper bound on relic density and safe against
Landau pole. All three benchmark points satisfy the LUX constraint.
However, as discussed earlier, the constraint from muon (g-2) is not incorporated.
\begin{table}[t]
\begin{center}
  \begin{tabular}{|c||c|c||c|}
   \hline
     Input   & \multicolumn{2}{c||}{Bino-like LSP} & Singlino-like LSP \\ 
 \cline{2-4}
 Parameters  & BP1 & BP2 & BP3 \\
   \hline
 $\lambda$             & 0.70 & 0.70 & 0.69 \\
 $\kappa$              & 0.20 & 0.20 & 0.11 \\
 $A_{\lambda}$  (GeV)  & 1800 & 1730 & 1950 \\
 $A_{\kappa}$   (GeV)  & -135 & -135 & -135 \\
 $\mu_{_{eff}}$ (GeV)  &  900 &  900 &  900 \\
 $\tan\beta$           &    2 &    2 &    2 \\
 $A_t$          (GeV)  & -100 & -250 & -200 \\
       $M_1$ (GeV) & 294 & 294 & 600 \\
 $m_{\tilde{Q}_{3}}$ (GeV) & 794 & 805 & 800 \\
 $m_{\tilde{U}_{3}}$ (GeV) & 834 & 836 & 832 \\
 $m_{\tilde{D}_{3}}$ (GeV) & 3000 & 3000  & 3000 \\
  & & & \\
   \hline
   \hline
 Observables & BP1 & BP2 & BP3 \\
   \hline
  Bino/Singlino & 0.99 & 0.99 & 0.99 \\
   fraction in the LSP    & & & \\
   \hline
  $m_h$ (GeV) & 123.3  &  125.0  &  123.0 \\
 $\msttwo$ (GeV) & 480.8 & 518.6 & 500.4 \\
 $\msbone$ (GeV) & 372.8 & 404.0 & 387.9 \\
 $\mstone$ (GeV) & 320.5 & 312.1 & 304.2 \\
 $m_{_{LSP}}$ (GeV) & 290.4 & 290.4 & 278.8 \\
   \hline
 BR($\sttwo \to \stone Z$)            & 0.57 & 0.67 & 0.67 \\
 BR($\sttwo \to \sbone W^+$)          & 0.39 & 0.30 & 0.32 \\
 BR($\sbone \to \stone W^-/W^{-*}$)   & 0.05 & 0.90 & $\sim 1$ \\
 BR($\sbone \to b \;\; \mathrm{LSP}$) & 0.95 & 0.10 & $\sim 0$ \\
   \hline
$\Gamma_{\stone}$ (GeV)  &  $3.4 \times 10^{-12}$ & $5.2 \times 10^{-13}$
                         &  $4.8 \times 10^{-14}$ \\
 BR[$\stone \to c \ntrlone$]  &  0.15  &  0.52  &  0.99 \\
 BR[$\stone \to b f \bar{f}' \ntrlone$]  &  0.85  &  0.48  &  0.01 \\
   \hline
 $\sigma(pp \to \sttwo \sttwos)_{_{\mathrm{LHC13}}}$ (fb)  & 633 & 412 & 503 \\
 $\sigma(pp \to \sttwo \sttwos)_{_{\mathrm{LHC14}}}$ (fb)  & 804 & 523 & 641 \\
   \hline
  \end{tabular}
\end{center}
\caption{Benchmark set of input parameters for the cases with bino-dominated 
(BP1 and BP2) and the singlino-dominated (BP3) LSP along with the resulting 
spectra and the branching fractions for the important decay modes
including those for the lighter stop (see figure \ref{fig:branchings}).
The total decay width of the lighter stop (see figure \ref{fig:widths}) 
and the production cross sections of the $\sttwo$-pair (at the NLO+NLL accuracy, 
for 13 TeV (LHC-13) and 14 TeV (LHC-14) LHC runs; 
see figure \ref{fig:xsec}) are also indicated.
Soft masses of the squarks from the first two generations, those for all 
the sleptons and the $SU(3)$ gaugino mass, $M_3$ are set to 3 TeV. The
$SU(2)$ gaugino mass $M_2$ is set to 1.2 TeV.
The SM-like Higgs boson happens to be the lightest Higgs boson for all the
three benchmark points. See text for details.}
\label{tab:benchmarks}
\end{table}

The two (pure) bino-like LSP scenarios, BP1 and BP2 (with 99\% bino-admixture), 
differ only in the values of the soft mass parameters for the $U(1)$ gaugino 
and those for the doublet and the up-type singlet squarks from the third generation 
(i.e., $m_{\tilde{Q}_{3}}$ and $m_{\tilde{U}_{3}}$). Slightly different values for the
trilinear soft SUSY breaking term $A_t$ have been chosen. The purpose of
presenting two such benchmark points in the bino-dominated LSP scenario is
just to demonstrate that significantly different branching fractions in the
cascades are attainable which would have drastic bearings for the final states. 
Such variations could also be achieved by varying some 
of the NMSSM inputs simultaneously and hence the specific choices of the
input parameters need not be over-stressed. Nonetheless, some 
aspects of these choices, like largeness of $\lambda$ and an appropriate
intermediate value for $\kappa$ (thus, together, ensuring constraints 
pertaining to non-appearance of Landau pole up to the unification scale),
not so large values of the soft masses $m_{\tilde{Q}_{3}}$ and $m_{\tilde{U}_{3}}$ are
very much of a defining nature for the scenario under consideration.
It is observed that not too different values of $m_{\tilde{Q}_{3}}$ and $m_{\tilde{U}_{3}}$
can be afforded since the scenario we are considering is intrinsically of a 
`compressed' nature. Naturally, the relevant spectra involving the stop, 
sbottom and the LSP are not so different. Indeed, it is 
interesting to note that the decay branching fractions of the stop and the 
sbottom squarks to the available modes are markedly different for BP1 and BP2.
As can be expected (and to be exploited later in this work), these differences 
would inevitably show up in the strengths of different possible final states 
through which such scenarios can be probed. The bottom-line is that
the search channels can be rather sensitive to the actual spectrum in such
compressed scenarios and this can be expected to be a somewhat generic feature. 

BP3 presents a (an almost pure) singlino-like LSP scenario.
Singlino-domination is ensured by a relatively low value of $\kappa$ and
with a large value of $M_1$. In this case, the second lightest neutralino
state becomes bino-dominated as long as $\mu_{eff}$ and $M_2$ are much larger
than $M_1$, which is the case for BP3. Note that the choice $M_1=600$ GeV
makes $m_{\chi_2^0}$ minimally large ($\sim M_1=600$ GeV) that ensures the 
setup we like to adhere to in the sense that a new decay mode like 
$\sttwo \to t \chi_2^0$ does not open up. Hence such a choice would not
broadly affect the collider phenomenology we address in this work while 
a richer phenomenology could be envisaged with a lighter bino-like neutralino
on its own right.

Note that the spectrum for BP3 is not again very different from the 
bino-dominated cases (BP1 and BP2). In all the three cases, $\msttwo \sim 500$
GeV, $\msbone \sim 400$ GeV, $\mstone \gtrsim 300$ GeV and $m_{LSP} \sim 270-300$ 
GeV. The essential difference between the scenario of BP3 and those of BP1 and BP2
is that in BP3 $\sbone$ can only (for all practical purposes) decay to states
involving an on- or off-shell $W$-boson. This is since for BP3, singlino-domination 
in the LSP efficiently suppresses the coupling 
$\sbone-b-\ntrlone$ to an insignificant level. In fact, in the singlino limit,
such a suppression is generic and this affects the decay of the NLSP 
(next to LSP) stop 
($\stone$) as well. However, the singlino-like LSP state being the only state which
$\stone$ could decay to, the pertinent issue is to what extent its decay width 
is affected by the suppressed coupling. As we will discuss later, this could
clearly have implications for the LHC. The important branching fractions for
the involved states are presented at the end of table \ref{tab:benchmarks}
for the three benchmark points. The extent of contrast in these branching fractions
is apparent. These are instrumental in shaping up the characteristic final states
for each of these scenarios through which they can be probed at the LHC.

Further, it is observed that only low values of $\tan\beta$ are consistent 
with the scenario we consider that requires larger $\lambda$. This corroborates 
the findings of reference \cite{King:2014xwa} with the exception that in our 
case only the lightest Higgs state can be the SM-like one. This fact is intimately
related to our choice of the LSP mass ($\sim 300$ GeV); be it a bino-like or a
singlino-like LSP, and that of $\mu_{eff}$. This can be roughly
understood as follows. 
As discussed in the beginning of section \ref{subsec:sm-higgs}, the squared mass 
of the SM-like (CP-even) Higgs state is given by \ref{eq:hmass}
while the same for the singlet-like 
CP-even Higgs boson is of the form $m_{ss}^2 = \kappa v_S(A_{\kappa}+4 \kappa v_S)$
\cite{Ellwanger:2011sk, Ellwanger:2009dp}.  
Note that the entry for the singlino mass as shown in the neutralino mass matrix,
given by equation \ref{mneut}, is ${\cal M}_0^{55} = 2 \kappa v_S$.
The product $\kappa v_S$ thus appears in both $m_{ss}^2$ and ${\cal M}_0^{55}$.
A bino-like LSP with a mass $\sim 300$ GeV can now be obtained by 
making the mass of the singlino-like neutralino to be much larger, 
i.e., by requiring $2 \kappa v_S >> 300$ GeV. This in
turn makes the mass of the singlet-like CP-even Higgs boson ($m_{ss}$) heavier
than the SM-like Higgs boson thus making the latter the lightest Higgs state.
On the other hand, for a singlino-like neutralino with mass around 300 GeV,
$\kappa v_S$ should have a low to moderate value and it is a priori not impossible 
to have the lightest Higgs boson to be singlet-like and the SM-like Higgs boson 
becoming the second lightest Higgs boson. However, our requirement of a somewhat 
large $\mu_{eff}=\lambda v_S$ ($>\msttwo$) ensures a large $\kappa v_S$ 
(for a given $\lambda \lesssim 1$) thus lifting up $m_{ss}$. This renders the
singlet-like Higgs boson heavier. Hence again the SM-like Higgs boson turns out
to be the lightest Higgs state.
\begin{figure}[ht!]
\centering
\includegraphics[height=85mm, width=\textwidth]{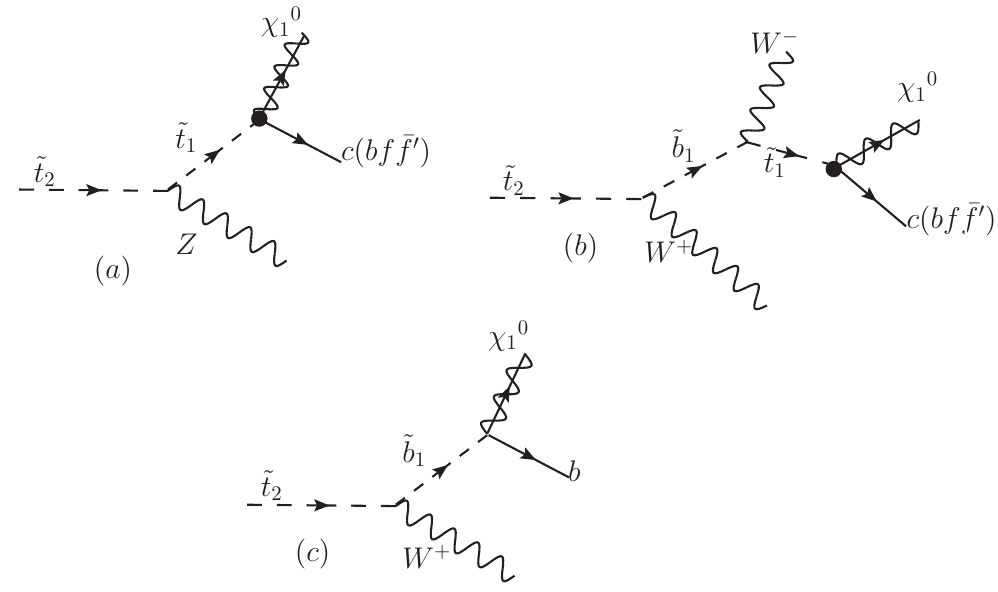}
\caption{Possible cascades of $\sttwo$ in the simplified scenario we consider.
Competing two-body (to $c \ntrlone$) and four-body 
(to $b f \bar{f}' \ntrlone$) decays of $\stone$ are also considered for which
the effective vertices are presented as black filled blobs.}
\label{fig:feyndiag}
\end{figure}
%
\section{A pair of light stops at the LHC}
%
In this section, we present the phenomenology of a pair of light stops
produced directly at the LHC by studying the cascades they 
undergo in a `simplified' NMSSM scenario described earlier. We adhere 
to the broad scenario of figure \ref{fig:cascade}. All three benchmark 
points discussed in section \ref{subsec:benchmarks} conform to this. 
There the choice of input parameters respects the latest experimental 
bounds. However, as already clarified in section \ref{subsec:benchmarks}, 
we do not subject these points to the observed value of muon ($g-2$) 
and those pertaining to the DM sector. 
%
\subsection{Cascade decay of the heavier stop and possible final states}
\label{subsec:cascade-stop2}
%
At the LHC, characteristic signatures of such a scenario with two light stops 
are triggered by the production and decay of the heavier stop
($\sttwo$). The lighter stop, being the NLSP and thus could decay only in 
a very restricted way, essentially plays a supporting role. The strong 
production cross section (which by far dominates) of the pair $\sttwo \sttwos$ 
is a simple function of $\msttwo$. Once produced, $\sttwo$ could cascade to the 
LSP in three possible ways as shown in figure \ref{fig:feyndiag}. Figures
\ref{fig:feyndiag}(a) and \ref{fig:feyndiag}(b) also indicate that $\stone$
could have FV and 4B decays. These could potentially
compete in our present scenario. We would briefly discuss the phenomenological 
implications of such possibilities later in this subsection.
For convenience, we show in table \ref{tab:stop-decay} the final states that
each of these branches leads to. We denote a pair of leptons originating in
the decay of a $Z$-boson by a brace underneath. The forward slashes separate
the possible alternate final states that the decays of $Z$- or a $W$-boson(s) 
(appearing in a cascade) may result in. Thus, the final states presented in
the third column of table \ref{tab:stop-decay} serve as the seeds when
constructing the overall final states arising from cascades of a pair of 
$\sttwo \sttwos$.
\begin{table}[ht!]
\begin{center}
{\footnotesize
  \begin{tabular}{ |c|c|c| }
  \hline
 Diagram  &  Cascade modes of $\sttwo$ &   Cascade products  \\  \hline   
            & & \\
    $(a)$ & $\sttwo \to \stone Z$, $\quad$ $\stone \to (c/b f\bar{f}^\prime) \, \ntrlone$  
  & ($\underbrace{\ell ~ \bar{\ell}} ~/~ 2j) + 1c/1b + \slashed{E}_T $    \\  
            & & \\
    $(b)$ & $\sttwo \to \sbone \wplus$, $\quad$ $\sbone \to \stone \wminus$,
    $\quad$ $\stone \to (c/ b f\bar{f}^\prime) \, \ntrlone$   
          & ($\ell ~\bar{\ell}$ $\, / \,$ $1~\ell(\bar{\ell})+2j$ $\, / \,$ $4j$) $+ 1c/1b + \slashed{E}_T $ \\  
            & & \\
    $(c)$ & $\sttwo \to \sbone \wplus$, $\quad$ $\sbone \to b \ntrlone$   & ($1 \ell ~ /~ 2j) + 1 b + \slashed{E}_T $\\
            & (only if the LSP is not singlino-like) & \\
    \hline
  \end{tabular}
}
\end{center}
\caption{Possible decay chains of $\tilde{t}_2$ following the cascade 
diagrams shown in figure \ref{fig:feyndiag}. The third column presents the
possible final states arising from the respective cascades. For the decay of
$\sttwos$, each entry needs to be charge-conjugated. 
}
\label{tab:stop-decay}
\end{table}      

In table \ref{tab:fstates} we display the different finals states constructed
out of possible (6 in all) cascade-combinations of $\sttwo$ and $\sttwos$. 
The combinations of individual cascades ($a$, $b$ and $c$) in the first column 
refer directly to the diagrams/cascades presented in figure 
\ref{fig:feyndiag}/table \ref{tab:stop-decay}.
For clarity, we divide the final states arising from each combination of 
cascades presented in the first column into two parts: (i) states appearing in 
the decays of $\stone/\sbone$ in the last phase of the SUSY cascade that
include jets comprising of bottom, charm and other light quarks, leptons 
along with missing energy carriers like the neutrinos and the LSP, 
(ii) the states (jets, leptons and neutrinos) arising in the decays 
of $Z$- and/or $W$-bosons appearing in the cascade decays of $\sttwo$.
As mentioned earlier, braces under a lepton pair
indicate that they are coming from the decay of a $Z$-boson. Note that
although there is a multitude of possible final states, these
are all at the parton-level. In our actual collider simulation which we
describe in the next subsection, the multi-jet final states are treated to a
very good degree of sophistication by using advanced jet-related techniques. 
These render only final states with limited jet multiplicities to be of 
practical interest.

\begin{table}[h]
\begin{center}
{\footnotesize
  \begin{tabular}{ |c|c|c|c| }
    \hline
       Cascade    &  On-shell &  \multicolumn{2}{c|}{\hskip -50pt Final states arising 
                                         from decays of a
                                         $\sttwo \sttwos$ pair} \\ 
                   \cline{3-4}
     combination  &  Gauge Bosons  & From $\stone$ decay & From $W^\pm$ \& $Z$ decays
     (in $\sttwo$, $\sbone$ cascades) \\
    \hline
       &    &  & \\
    $(a) ~ (a)^{*}$ &  $ZZ$ & $\left.\begin{tabular}{@{\ }r@{}}   
   $2c + \ntrlone \ntrlone$ $\; {\mathbf{(A)}}$ \\
   $2b + \ell^+ \ell^- \, \nu \bar{\nu} + \ntrlone \ntrlone$  $\; {\mathbf{(B)}}$  \\
   $2b + 4j + \ntrlone \ntrlone$    $\; {\mathbf{(C)}}$\\
   $1c + 1b +  \ell \, \nu  + \ntrlone \ntrlone$   $\; {\mathbf{(D)}}$ \\
   $1c + 1b +  jj  + \ntrlone \ntrlone$    $\; {\mathbf{(E)}}$ \end{tabular} \right \}$ 
 & $ 2\underbrace{\bar{\ell} \ell} $, ~$ \underbrace{\bar{\ell} \ell}+2j $, ~ $ \underbrace{\bar{\ell} \ell} $ \\ 
 & & &  \\
%
  $ (b) ~ (b)^{*}$ & $2(W^+W^-)$ &  Same as above & $2\bar{\ell}+2 \ell$, \quad 
                            $ 2\bar{\ell}(\ell)+\ell(\bar{\ell}) + 2j$, \\
   & & &     ~$ \bar{\ell} \ell +4j $, \quad $\mathrm{SSDL}+4j$ \\ 
 & & &  \\
    $ (a) ~ (b)^* + h.c.$  &  $Z(W^+W^-)$  &  Same as  above  &  
        $\underbrace{\bar{\ell} \ell} + \bar{\ell} \ell $, 
  \quad $ \underbrace{\bar{\ell} \ell}+\bar{\ell}(\ell)+2j $, \\
 & & &  $\bar{\ell} \ell$,~ $ \bar{\ell} \ell +2j $, 
  \quad $ \underbrace{\bar{\ell} \ell} + 4j $\\ 
 & & &  \\
    $ (c) ~(c)^{*} $ & $W^+W^-$  & {\bf B} and {\bf C}  &  $ \bar{\ell} \ell $ \\ 
 & & &  \\
    $ (a) ~ (c)^* + h.c.$ &  $ZW^\pm$  & {\bf B}, {\bf C}, {\bf D} and {\bf E}  &   $\underbrace{\bar{\ell} \ell}+\bar{\ell} (\ell) $,~ $ \underbrace{\bar{\ell} \ell} + 2j $ \\
 & & &  \\
    $ (b) ~ (c)^* + h.c.$ &  $(W^+W^-)W^\pm$  &  Same as above  &  
    $2\bar{\ell}(\ell) + \ell(\bar{\ell}) $, 
$\mathrm{SSDL} + 2j $,~$ \bar{\ell} \ell+ 2j $  \\
 \hline
 \hline
\end{tabular}
}
\end{center}
\caption{All possible final states from the decays of a pair of 
$\tilde{t}_2$~$\tilde{t}_2^*$. 
Each row corresponds to specific cascades of 
$\tilde{t}_2$ and ~$\tilde{t}_2^*$. For each row, the set of particles
appearing in the final state combines the ones from the third column
(from the decays of $\stone / \stones$) and those from the fourth column
(coming from the decays of $Z$- and $W$-bosons produced in the cascades).
Only those final states with $\geq 2$ leptons are shown. 
Presence of neutrinos is implied in leptonic decays of $W^\pm$ and when
$Z$ decays invisibly.
}
\label{tab:fstates}
\end{table}

Before we end this subsection, a brief discussion over the possible decays of
$\stone$ would be in place. As pointed out already, $\stone$ could undergo the
two-body FV decay to $c \ntrlone$ and the 4B decay to $b f \bar{f}^\prime \ntrlone$.
Recently, the implications of the competition between these decay modes have been
much appreciated \cite{Krizka:2012ah, Delgado:2012eu, Belanger:2013oka, 
Grober:2014aha, Ferretti:2015dea}, 
in particular, in the context of compressed scenarios in the MSSM with a light
$\stone$.\footnote{Various strategies to probe the stops squarks in difficult situations 
with compressed spectrum are proposed in the literature (see, for example,
references \cite{Dutta:2013gga, Cho:2014yma, Eckel:2014wta} and references therein).} 
These are subsequently followed up by the LHC collaborations 
\cite{Aad:2014nra, Aad:2014kra}.

Our concern is over a possibly longer lifetime of $\stone$ which would have
crucial effect on its phenomenology and on the consequent experimental strategy 
to find them. In our scenario, this may happen for two different reasons. First, 
for an NLSP $\stone$ with a rather small mass-split with the LSP, there is a 
looming possibility that its decay width becomes critically suppressed by the 
phase space. Second, in the NMSSM, for a singlino-like LSP, the decay width of 
$\stone$ can be severely affected due to its suppressed (effective) couplings to 
the LSP. Worse could be the situation when these two issues work in tandem. 
In reference \cite{Covi:2014fba}, phenomenology with non-prompt decays of (long-lived) 
NSLP $\stone$ has been discussed in much details where $\stone$ is either stable 
across the pixel or tracker detector resulting in a displaced vertex or stable over 
the whole detector dimension (metastable) thus leaving a charge-track. It is 
important to note that for a metastable $\stone$ the latest LHC analyses already
exclude $\mstone < 900$ GeV \cite{ATLAS:2014fka} (except for a possible caveat
recently pointed out in reference \cite{delaPuente:2015vja}). 
Hence any analysis with light $\stone$ must
ensure that this does not become long-lived.
\begin{figure}[t]
\centering
\includegraphics[height=0.22\textheight, width=0.49\columnwidth , clip]{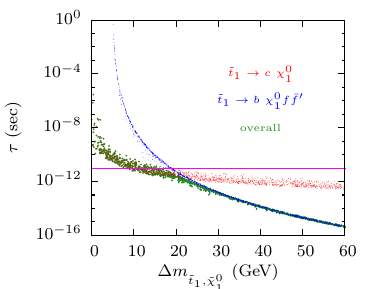}
\includegraphics[height=0.22\textheight, width=0.49\columnwidth , clip]{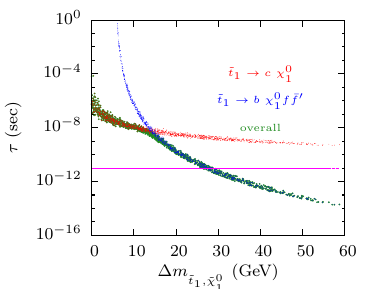}
\caption{Decay life-time of $\stone$ (at rest) if only the two-body $c\ntrlone$ mode 
(in red) or only the four-body $b f \bar{f}' \ntrlone$ mode (in blue) were possible
as functions of the mass-split ($\Delta{m}_{\stone, \ntrlone}$) between $\stone$
and $\ntrlone$. Points in green indicate the values of the actual lifetime of $\stone$ 
(at rest) and result from summing up its decay widths in the two modes. The left plot
is for the case of bino-dominated LSP while the right one stands for the case of the
singlino-dominated LSP. The scattered points are obtained by scanning over the NMSSM 
parameter space. See text for details.}
\label{fig:widths}
\end{figure}

In figure \ref{fig:widths} we show via scatter plots the individual lifetimes of 
$\stone$ with only FV 
(in red), only 4B (in blue) decays and the overall lifetime (in green). These are 
shown for both the bino-dominated LSP (left) and the singlino-dominated LSP 
(right) cases. Note that while the FV decay width is calculated by the 
{\tt NMSSMTools}, the 4B ones are not. Hence we use {\tt MadGraph5\_aMC@NLO v2.1.2} 
\cite{Alwall:2014hca} to calculate the latter
and then get them cross-checked in {\tt CalcHEP (v3.6.23)} \cite{Belyaev:2012qa}. 
For the present purpose, the scans are done around the
respective benchmark points with $|A_t| \leq 300$ GeV and 
$700 \, \mathrm{GeV} \leq m_{\tilde{Q}_3}, m_{\tilde{U}_3} \leq 1 \,\mathrm{TeV}$, 
keeping all remaining parameters fixed.
For larger mass-splits ($\Delta m_{\stone, \ntrlone}$) between 
$\stone$ and the LSP, the four-body decay dominates. Hence the green bands overlap 
with the blue bands. For smaller values of $\Delta m_{\stone, \ntrlone}$, the
bottom quark mass ($\sim 5$ GeV) stands in the way for the four-body decay to 
dominate. Thus, the FV decay becomes important and the green bands superpose on 
the red bands in this regime. The scattered points are the results of scans 
over the NMSSM parameter space. The horizontal lines in the two plots represent 
a conservatively chosen critical lifetime of $10^{-11}$ second in the rest
frame of the decaying $\stone$\footnote{For simplicity, we skip the 
boost factor $\beta\gamma$ where the $\beta={v \over c}$ and 
$\gamma={(1-{v^2 \over c^2})^{-{1 \over 2}}}$. At the LHC, $\beta\gamma$ has a 
distribution and may result in an increase of the $\stone$ lifetime by a factor 
of 1.5 to 2 \cite{Covi:2014fba}.}. 
This corresponds to a decay-length (given by $\beta\gamma c \tau$)
of a few millimeters and can be considered to be prompt for a new physics excitation.
In any case, note that for $\Delta m_{\stone, \ntrlone} \gtrsim 25$ GeV, which 
is the case in our present analysis, the overall lifetime is smaller than this 
critical value. Thus, we are always in the regime where $\stone$ undergoes prompt
decays irrespective of whether the LSP is bino- or singlino-dominated. 

\begin{figure}[t]
\centering
\includegraphics[height=0.22\textheight, width=0.49\columnwidth , clip]{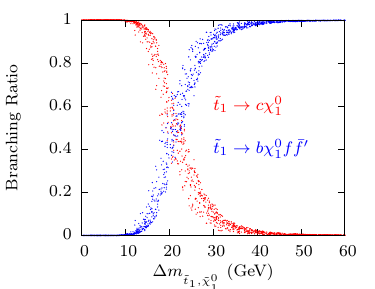}
\includegraphics[height=0.22\textheight, width=0.49\columnwidth , clip]{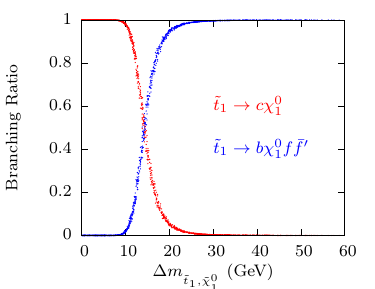}
\caption{Same as in the case of figure \ref{fig:widths} but presenting the respective
branching fractions of $\stone$.}
\label{fig:branchings}
\end{figure}

In figure \ref{fig:branchings} we illustrate the corresponding variations of 
the branching fractions in the FV and 4B modes for the bino- (left) and the 
singlino-dominated (right) LSP. Clearly, for $\Delta m_{\stone, \ntrlone} \gtrsim 20$ 
GeV which is what we have for our benchmark scenarios, the four-body decay takes over.
This substantiates the recent appreciation of the importance of the four-body 
decay of $\stone$. Furthermore, an early onset (for a smaller 
$\Delta m_{\stone, \ntrlone}$) of domination for the four-body branching fraction
does not escape notice. This is expected since the FV decay experiences 
simultaneous suppressions from the loop-process and the diminished coupling for 
a singlino-like LSP.

%
\subsection{Signatures at the LHC}
%
From table \ref{tab:fstates} it is clear that charm and/or bottom 
quark(s) in the final states are unavoidable. The sources of the bottom quarks 
are the two-body decays $\sbone \to b \ntrlone$ and the four-body decays of the $\stone$.
If the LSP is singlino-like, the former decay is disfavored and the cascades end
with decays of $\stone/\stones$.

Thus, identifying the $c$- and the $b$-\emph{jets} would definitely help 
probe the scenario \cite{atlas-bc-tag}. 
As for $c$-tagging, it is understood to be a dedicated 
technique with low to moderate efficiencies. This is true, in particular, for
charm quarks coming from the decays of the stops having a small mass-split
with the LSP. Therefore, we would not demand tagging of $c$-\emph{jet(s)}, 
although, if becomes possible, this could be rather 
helpful\footnote{A recent ATLAS Note \cite{atlas-c-tag} 
discusses a dedicated $c$-tagging algorithm (in reference to the process 
$pp \to \stone \stones \to c \, \ntrlone \, c \ntrlone$).
The study reveals that a medium/conservative (loose) set of criteria yields 
on an average
a $\approx 20$\% ($\approx 95$\%) $c$-tagging efficiency. The medium (loose)
criterion seems to be optimal when the background is dominated by light- and 
$b$-quark jets ($t \bar{t}$ processes). Given the rather preliminary nature of 
the study and a more complex topology that we are considering for our signal, 
it makes sense that at this point we talk about the conservative values of 
the projected efficiency. For a rather soft $p_T^{charm} \lesssim 20$ GeV 
that we consider, the conservative efficiency now ranges between 10-20\% which 
is a little too small for our present purpose.}.
Note that although the scenario is of a compressed kind, in the present case,
the charm quarks are not coming from stops that are directly 
produced in $pp$ collisions. Rather, these are coming from the decays of heavier 
states like $\sttwo$ and/or $\sbone$. Thus, the $\stone\stones$ system would not be back-to-back (in the plane transverse
to the beam axis) unlike when these are directly produced. Such a production 
of $\stone\stones$ under cascades of heavier states, thus, to a limited extent, 
could play the role of a hard jet from the initial state radiation (ISR) against 
which the  $\stone\stones$ recoils in case of their direct production. This 
would help increase the amount of $\slashed{E}_T$ along with the $p_T$ of the 
$c$-\emph{jets} even without the presence of a hard ISR jet.
As for the $b$-\emph{jets}, a noteworthy aspect of such a scenario is their 
natural dearth in the final state. The depletion is primarily at the level of
their rates (effective branching fractions; see table \ref{tab:benchmarks}).
The depletion in their rates is severe when BR($\sbone \to b \ntrlone$) 
becomes further suppressed, typical to the case with a singlino-dominated LSP. 
However, to be convinced of their presence, we incorporate $b$-tagging 
in our analysis.

On the other hand, with as many as three possible SM gauge bosons 
($Z$ and $W^\pm$) that
may appear in the cascade of a $\sttwo$-pair, lepton-rich final states would
be common. These leptons could naturally serve as the cleaner probes to such 
a scenario at the future runs of the LHC. Finding the footprints of 
one or more on-shell $Z$-bosons (via their reconstructions from pairs of
OSSF leptons)
in conjunction with finding leptons with origins in decays of on-shell
$W$-bosons hold the key. Hence we confine our analysis to final
states having three to four leptons.
At this point, the following few issues pertaining to the scenario
under consideration may be noted down:
\begin{itemize}
\item when no $b$-\emph{jet} is available in the final state, finding
      one or more $c$-\emph{jets} could be helpful. 
      There are three ways in which $b$-\emph{jets} could get depleted 
      in our scenario\footnote{It is worthwhile to note that
      even when we deal with three squarks from the third generation,
      there are various different possibilities which result in
      final states with depleted or even zero $b$-\emph{jets}.
      This can be traced back to the fact that in our scenario
      the lighter chargino is heavier than $\sttwo$ and hence the decay
      $\sttwo \to b \tilde{\chi}^+_1$ is closed. Thus, 
      out of the three light squarks, only the decay of $\sbone$ could
      lead to a $b$-\emph{jet}.
}
\begin{itemize}
\item when the LSP is singlino-like and hence $\sbone \to b \ntrlone$ 
      is negligible,
\item when the LSP is bino-like but $\sttwo$ dominantly decays to
      $\stone Z$ followed by $\stone \to c \ntrlone$ and
\item when the LSP is again bino-dominated but the decay $
      \sbone \to \stone W^-$ dominates over the decay $\sbone \to b \ntrlone$. 
\end{itemize}
\item $b$-\emph{jets} always become handy whenever they appear (only favored for
      a bino-dominated LSP). Also, for our benchmark points the splitting 
      $m_{\sbone} - \mntrlone$ is around 100 GeV thus making the $b$-tagging 
      efficiency near-maximal for those $b$-s that come from the decays of 
      $\sbone$. The $b$-quarks having origins in the four-body decays of 
      $\stone/\stones$ are presumably too soft to be efficiently detected in the 
      present setup.
\item A naturally `charm-less' final state is possible (see the fourth entry 
      ($(c) (c^*)$) of table \ref{tab:fstates}) when tagging of the 
      $b$-\emph{jets} would be all the more important. The $\sttwo$-cascade that leads
      to this channel (see figure \ref{fig:feyndiag}{\color{blue}c}) involves a
      $W$-boson. Thus, there can be up to two leptons in a charm-less final 
      state.
\item Charm-less final states are also possible when both $\stone$ 
      states decay via the four-body mode (see, for example, the first entry 
      of table \ref{tab:fstates} 
      (mode {\bf B})). Leptons that are presumably too soft 
      to be easily detected may appear in such decays.
\item It may also be noted that such four-body decays of $\stone/\stones$ could
      result in $c$-quarks (via off-shell $W$-bosons leading to the `$c s$' final 
      state (see entry $((a)(a^*))$ of table \ref{tab:fstates}, mode {\bf C}), which 
      might have an even softer $p_T$ spectrum when compared to the same for the 
      $c$-quark arising from the decay $\stone \to c \ntrlone$.
\item If detectable, final states with various possible combinations of $c$- and 
      $b$-\emph{jet} multiplicities could turn out to be efficient and corroborative
      probes for an in-depth understanding of such a scenario.
\end{itemize}

To define the optimal signal regions, following the above discussions,
we first pick up five appropriate final states.
These are listed in table \ref{tab:signals}. The second column presents 
the actual search modes
and these can be directly traced back to table \ref{tab:fstates} and thus,
in turn, to the contributing cascades of $\sttwo$ illustrated in
figure \ref{fig:feyndiag}. 
In the last column of table \ref{tab:signals} we indicate the sources of dominant
SM backgrounds against each of the signal final states. It is to be noted
that bare $t\bar{t}$ production, by itself, is not a dominant background for the 
multilepton ($n_\ell \geq 3)$ final states under consideration except for the 
SSDL one.
The choice of these final states are prompted by their cleanliness due to the
presence of three or more leptons. These leptons have their origins in the $Z$- and the $W$-bosons appearing in the cascades of $\sttwo$. Hence at least one 
pair of opposite sign same flavor (OSSF) (with a maximum of two such pairs) 
can be reconstructed to the mass of the $Z$-boson. 
\begin{table}[ht!]
\begin{center}
  \begin{tabular}{ |c|c|c| }
  \hline
 Channel &   Search channel & Dominant \\  
    ID   &   & backgrounds \\
\hline   
             & & \\
    $\mathrm{SRSSDL0b}$    & $\mathrm{SSDL} + (4,6)j + 0 b+ \slashed{E}_T$
          & $ t\bar{t}$, $t\bar{t}W$, $t\bar{t}Z$, $ZW$ \\
             & & \\
    $\mathrm{SRSSDL1b}$    & $\mathrm{SSDL} + (3,4)j (1b$-$jet) + \slashed{E}_T$
          & $ t\bar{t}$, $t\bar{t}W$, $t\bar{t}Z$, $ZW$ \\
             & & \\
    $\mathrm{SR3\ell 0b}$  & $3 \ell + (3,4)j + 0 b+ \slashed{E}_T$  
          & $ZW$, $t\bar{t}$, $t\bar{t}Z$, $ZZW$ \\
             & & \\
    $\mathrm{SR3\ell1b}$   & $3 \ell + \leq 2 j \; (1 b$-$jet) + \slashed{E}_T$ 
                           & $t\bar{t}Z$, $t\bar{t}$, $t\bar{t}W$, $ZW$    \\
             & & \\
    $\mathrm{SR4\ell0b}$   & $4 \ell + (1,2)j + 0 b+ \slashed{E}_T$
          & $ZZ$, $ZWW$, $ZZW$ \\
             & & \\
    \hline
  \end{tabular}
\end{center}
\caption{Definition of the signal regions in terms of the actual search 
channels undertaken in the present analysis. Leptons have their origins 
in the $Z$- and the $W$-bosons
appearing in the cascades of $\sttwo$. At least one pair of leptons 
(opposite sign, same flavor (OSSF); electrons and muons only) is to come 
from an on-shell $Z$-boson. ``SSDL'' in the first two rows stands for same-sign 
dilepton final state. The last column presents the dominant SM background
processes corresponding to each final state. These are inclusive of two 
hard jets except for the $ZW$ and $t \bar{t}$ processes for which 
three-jet inclusive samples are used. 
}
\label{tab:signals}
\end{table}      

An exception to this is the final state with a lower lepton multiplicity 
in the form of same-sign dileptons (SSDL). This final state is traditionally 
known to be extremely clean given that the SM background is naturally 
suppressed. In our scenario, such a final state ($\mathrm{SRSSDL0b}$) is a possibility 
when $\sttwo \sttwos$ pair undergoes the cascade indicated by figure
\ref{fig:feyndiag}{\color{blue}(b)} (and the corresponding entry in table
\ref{tab:stop-decay}). More specifically, an SSDL final state arises from 
the leptonic decays of those $W$-bosons having the same electric charge 
that appear at two different stages of the cascades of $\sttwo$ and 
$\sttwos$.  

Note that in the scenario we consider, the charm and the bottom quark jets coming
from the decays of $\stone/\stones$ would be much softer
with $p_T^{jet} \lesssim 50$ GeV. Hence there is a fair
chance of losing such reconstructed jets at the detectors. Furthermore, as 
pointed out earlier, the final states are not generally rich in harder
$b$-\emph{jets}. In our analysis, we tag the $b$-\emph{jets} and accept or 
veto them as per requirement. As we will discuss shortly, the background to 
final state(s) with a $b$-\emph{jet} mainly arises from 
the generic processes like $t\bar{t}+jets$. However, the requirement of a 
large lepton-multiplicity effectively reduces this background. On the 
other hand, the final states without a $b$-\emph{jet}, in addition, draw backgrounds 
from the SM processes like vector-boson pairs plus extra jets. Demanding 
a minimum $\slashed{E}_T$ could help tame these backgrounds. 
As for the SM backgrounds, we consider three-jet inclusive samples for the $ZW$ 
(which can give rise to trileptons) and $t \bar{t}$ processes. For other 
background processes, we only consider samples inclusive of two extra hard jets.
\subsection{Simulation}
The exclusive signal process considered for our analysis is the basic 
production of a pair of $\sttwo$ at the LHC ($p p \to \sttwo \sttwos$)
followed by their cascades. The background processes we consider are already
listed in the previous subsection.
For both signal and the background processes, event samples are generated 
at the lowest order (LO) in perturbation theory using 
{\tt MadGraph5\_aMC@NLO v2.1.2} \cite{Alwall:2014hca}. {\tt CTEQ6L1} 
\cite{Pumplin:2002vw} parton distribution function is used with the 
factorization/renormalization scale set at the default MadGraph setting 
(i.e., at $m_{T_{\sttwo}} = \sqrt{\msttwo^2+p_{T_{\sttwo}}^2}$).
\begin{figure}[t]
\centering
\includegraphics[height=0.3\textheight, width=0.49\columnwidth , clip]{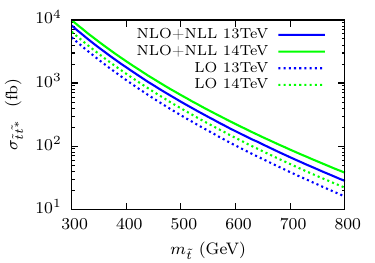}
\caption{Production cross sections (in femtobarn) of a pair of stops as functions
of the stop mass at LHC-13 and LHC-14. The lower two curves 
correspond to the LO estimations while the upper ones include NLO+NLL effects.}
\label{fig:xsec}
\end{figure}

For the signal, the next-to-leading-order (NLO) plus next-to-leading-log (NLL) 
cross sections are computed with the package {\tt NLL-fast (v3.0)} \cite{Beenakker:1997ut} 
which in turn uses the package {\tt Prospino2 (v2.1)} \cite{prospino2} for the NLO result. 
CTEQ6.6M parton distribution parametrization, default to the package, has been
employed for the purpose.
The NLO+NLL cross sections are later used to normalize the signal 
rate. In figure \ref{fig:xsec} we illustrate the variation of the 
$\sttwo \sttwos$ production cross section at LHC-13 and LHC-14
without and with the NLO+NLL correction as functions of $\sttwo$ mass.
On the other hand, to be on the conservative side, we apply $K$-factors
of 1.6 for the backgrounds from $t\bar{t}$ processes (with inclusive jets) 
and 1.3 for all the others.

Events generated for both the signal and the backgrounds are then showered 
and the unstable particles are decayed using {\tt Pythia (v6.426)}
\cite{Sjostrand:2006za} embedded within the Madgraph environment. 
For the signal, the decay branching fractions of various SUSY excitations
calculated within the framework of {\tt NMSSMTools} \cite{Ellwanger:2006rn} 
are provided to Pythia through the {\tt SLHA2} \cite{Allanach:2008qq} 
interface.
\footnote{The four-body decay width obtained from Madgraph (see section
\ref{subsec:cascade-stop2}) is inserted appropriately into the SLHA2 file generated 
by {\tt NMSSMTools} which contain all other decay information. This ensures 
correct estimation of $\stone$ branching fractions.}  
To avoid double counting 
of events in the background samples in the presence of extra hard partonic 
jets and the parton shower, we employ the MLM matching scheme 
\cite{Mangano:2006rw, Alwall:2007fs} with the 
variables {\tt xqcut} and {\tt qcut} set at appropriate values. 

\subsubsection{Reconstructing the physics objects}
The fast detector simulation framework {\tt DELPHES (v3.1.2)} 
\cite{deFavereau:2013fsa} is used to reconstruct final physics objects. 
The following steps are taken to obtain them in an ATLAS environment 
(in terms of the detector coverage).
\begin{itemize}
\item The jet-finding package {\tt FastJet (v3.0.6)} \cite{Cacciari:2011ma} 
embedded in {\tt DELPHES} is used to find the jets. The anti-$k_T$ jet 
algorithm is employed with the cone size set at 0.4, requiring a minimum 
$p_T^{jet}$ of 20 GeV and the pseudorapidity in the range 
$|\eta_{jet}| < 2.5$. A flat $b$-tag efficiency of 70\% is incorporated.
Furthermore, we consider a 20\% probability of a $c$-\emph{jet}
being mistagged as a $b$-\emph{jet} while the same with the light 
quark flavors are taken to be 1\% \cite{Aad:2014pda, atlas-bc-tag}.
\item Leptons are reconstructed with a minimum $p_T^{\ell epton}$ 
of 10 GeV and with $|\eta_{_{\ell epton}}| < 2.5$. For electrons and muons, we remove jets which lie 
within an angular distance $\Delta R \le 0.2$ from the lepton. Also, to 
increase the purity of electrons, it is required that the ratio of total 
$p_T$'s of the stray tracks within the cones of their identification to their 
own $p_T$'s is less that 0.1. The corresponding requirement for the muons is
that the maximum total $p_T$ of other tracks does not exceed 1.8 GeV.
\end{itemize}
The entire process of event-analysis is done within the modular and integrated 
framework of the popular package of {\tt MadAnalysis 5} ({\tt v1.1.11}) 
\cite{Conte:2012fm, Conte:2014zja} 
with the help of the in-built data analysis framework 
{\tt ROOT (v5.34.22)} \cite{cernroot} 
using the event format {\tt STDHEP} \cite{stdhep}.
\subsubsection{Event selection}
In table \ref{tab:cuts} we list three sets of kinematic cuts 
that are employed in our analysis for five exclusive final states: 
$\mathrm{SR3\ell 1b}$, $\mathrm{SR3\ell 0b}$, $\mathrm{SR4\ell0b}$,
$\mathrm{SRSSDL0b}$ and $\mathrm{SRSSDL1b}$. These cuts remain the same
for the corresponding (jet-)inclusive samples as well (which we discuss at
length) except for lifting the jet-related cuts 
(jet $p_T$, jet-multiplicity etc.). However,
these sets are common across all the three benchmark points that we consider
and for both LHC-13 and LHC-14. Graded cuts are applied on 
the $p_T$'s of leptons and jets and on the transverse masses 
(defined as 
$m_T= \sqrt{2 p_T^\ell \slashed{E}_T \left[ 1-\cos(\phi_\ell
     -\phi_{\slashed{E}_T} ) \right]}$)  
of the leptons based 
on their relative hardness. For this analysis, by leptons we mean electrons 
and muons. It may be noted that one cannot afford a more stringent lower
cut on $p_T$ of the jets. This is because, unlike in generic SUSY searches,
even involving squarks from the third generation, where jets may arise either
directly from rather heavy colored states or in the decays of heavy electroweak
gauginos, the sources of jets in the present case are only the $W/W^*$ bosons. 
On top of that, we work with not only light spectra but also somewhat 
compressed ones which further deplete the $p_T$ of the jets. 
\begin{table}[!htbp]
 \begin{center}
  \begin{tabular}{ ||c|c|c|c|c|c|| }
    \hline \hline
    Variables & $\mathrm{SR3\ell 1b}$ & $\mathrm{SR3\ell0b}$ & $\mathrm{SR4\ell0b}$ & $\mathrm{SRSSDL0b}$ & $\mathrm{SRSSDL1b}$ \\ \hline \hline
    $n_\ell$ & 3 & 3 & 4 & 2 (SSDL) & 2 (SSDL)\\ \hline
    $n_{\ell \ell}^{OSSF}$ & 1 & 1 & 2 & vetoed & vetoed \\ \hline 
    $M_{\ell \ell}^{OSSF}$ (GeV) & 
     \multicolumn{3}{c|}{$75  \leq M_{\ell \ell}^{OSSF} \leq 105$} 
    & -- & -- \\ \hline 
    $n_{jet}$ &  1 to 2 & 3 to 4 
   & 1 to 2 & 4 to 6 & 3 to 4 \\
   $n_{b{\mathrm -jet}}$ (incl.)
   & 1 & vetoed & vetoed & vetoed & 1 \\ \hline 
 $p_T^{j_{(n)}}$ (GeV)  & \multicolumn{5}{c||}{$p_T^{j_{(1,2,3,\geq 4)}}> (30, 30, 25, 20)$} \\
 $p_T^{b-jet}$ (GeV)  & \multicolumn{5}{r||}{$p_T^{b-jet} > 40$
(whenever applicable)} \\ \hline
 $p_T^\ell$ (GeV)  & \multicolumn{5}{c||}{$p_T^{\ell (1,2,3,4)} > (25,20,20,15)$} \\ \hline
 $m_T^{\ell_{(n)}}$ (GeV)  & \multicolumn{5}{c||}{$m_T^{\ell_{(1,2,3)}}> (140, 125, 100)$} \\ 
  & \multicolumn{5}{c||}{(not applicable for $\mathrm{SR4L0b}$)} \\ \hline
 $\slashed{E}_T$ (GeV)  & \multicolumn{5}{c||}{$\slashed{E}_T > 135$} \\ \hline \hline
 \end{tabular}
\caption{The complete definition of the signal regions (SR) denoting the final
states they represent and the respective sets of common selection cuts on
the physics objects that are independent of the benchmark scenarios and the 
LHC energies. 
By leptons only electron 
and muon are referred to. Other notations follow the standard conventions. 
}
\label{tab:cuts} 
 \end{center}
\end{table}   

\begin{figure}[htb]
\centering
\includegraphics[height=0.25\textheight, width=0.49\columnwidth , clip]{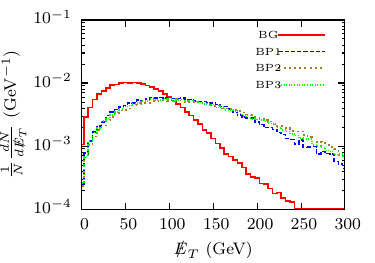}
\hskip -00pt
\includegraphics[height=0.25\textheight, width=0.49\columnwidth , clip]{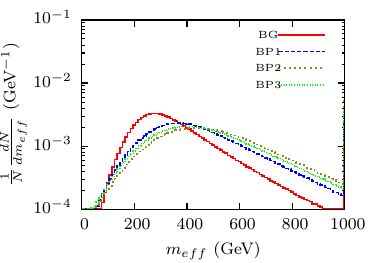}
\caption{$\slashed{E}_T$ distributions (left) and $m_{eff}$ distributions (right) 
for the SM background (in red) and for the signal in the inclusive multilepton final 
state for the three benchmark 
points (in blue, brown and green, respectively) and for LHC-13}
\label{fig:met-meff}
\end{figure}

For the same reason, one can expect only moderately hard $\slashed{E}_T$ for any 
of the targeted final states and for any of the benchmark points. This is apparent 
from figure \ref{fig:met-meff}. Therein we plot the $\slashed{E}_T$ distributions 
(left) and the effective mass ($m_{eff}=\sum_i p_{T_i}^{visible} + \slashed{E}_T$, `$i$'
standing for the $i$-th visible state;
right) for the SM background (in red) and for the three benchmark points (for 
inclusive final states) for LHC-13. The $\slashed{E}_T$ spectra 
peaks at around 100 GeV irrespective of the benchmark scenario. Note that this peak 
value is already on the smaller side given the mass of the LSP is around 300 GeV.
Consequently, one finds degraded peaks in the effective mass distributions at around 
400 GeV which characterize a compressed spectrum, while under normal circumstances, 
a flatter peak about the total mass ($\sim 1$ TeV) of the pair-produced particles 
($\sttwo$) is expected. We do not impose any separate cut on $m_{eff}$ as this does 
not improve the situation much once a cut on $\slashed{E}_T$ is imposed.

\begin{figure}[htb]
\centering
\includegraphics[width=\columnwidth , clip]{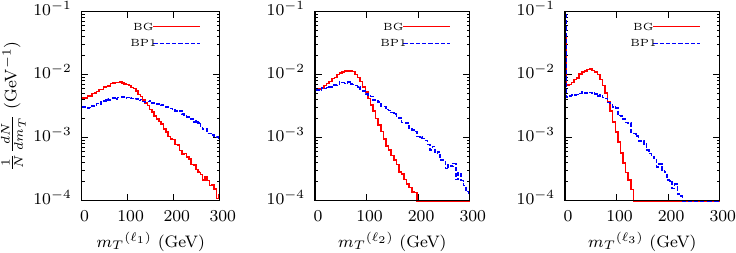}
\caption{$m_T$ distributions of the three hardest leptons $\ell_1$, $\ell_2$
and $\ell_3$ (in order of decreasing hardness; from left to right) for the 
SM background (in red) and for the signal benchmark point BP1 (in blue) 
at LHC-13 and in an inclusive sample. 
}
\label{fig:mt}
\end{figure}

In figure \ref{fig:mt}, from left to right, we illustrate the $m_T$ 
distributions of the three hardest leptons $\ell_1$, $\ell_2$ and $\ell_3$ 
with $p_T^{\ell (1)} > p_T^{\ell (2)} > p_T^{\ell (3)}$ for the 
SM background (in red) and for the signal benchmark point BP1 (in blue) 
in the inclusive multilepton final state at LHC-13.
We do not find any major discriminatory feature among the $m_T$
distributions of the leptons in different final states and for
LHC-14. Hence we stick to a common set of selection 
criteria for them as indicated in table \ref{tab:cuts}.
Note that for the 4-lepton final state ($\mathrm{SR4\ell0b}$), 
the leptons come from a pair of on-shell $Z$-bosons. 
Hence we do not impose the $m_T$-cuts for this final state.

For all the three variables we discussed, we do not separately present 
plots for LHC-14 as they are rather similar (except for, 
as expected, the tails getting extended slightly towards the harder side).
%
\subsubsection{Results and discussions}
%
In this subsection we discuss the results of our simulations for LHC-13 and
LHC-14, and for different signal
regions described in tables \ref{tab:signals} and \ref{tab:cuts}.
The purpose of the simulation is to understand how sensitive the
LHC experiments are to the proposed NMSSM scenario with two relatively 
light stops.

It is important to note that the data ($\sim 20$ fb$^{-1}$ for each of ATLAS and CMS) 
from LHC-8 are not at all sensitive to the scenario we 
propose. This is in spite of rather low stop masses with $\mstone \gtrsim 300$ 
GeV and $\msttwo \lesssim 500$ GeV that we choose in this 
work. This can be understood in the following way.
First, our scenario is somewhat of a compressed kind where the mass-split between 
the NLSP $\stone$ and the LSP ($\ntrlone$) does not exceed $\sim 30$ GeV. 
Thus, $\stone$ decays can decay only to the FV mode $c \ntrlone$  and the 
four-body $b f \bar{f}^\prime \ntrlone$ mode. We respect the relevant bound which is 
$\mstone \gtrsim 250$ GeV \cite{Aad:2014nra,cms-stop-1} and is
grossly applicable to our scenario (except under specific situations
when the LSP is singlino dominated and thus $\stone$ could become
relatively long-lived at the collider). Our choice of $\mstone \approx 300$ GeV 
is thus conservative to a good extent and it may be only natural to expect that 
LHC-8 would not be sensitive to such a stop squark.

However, the scenario is not all about the lighter stop on its
own. In fact, the combined phenomenology of the two stops
together is what defines the situation. Note that in terms of the
basic production mechanism, the phenomenology is entirely governed by
the pair-production rate of $\sttwo$. However, the final state 
may or may not involve a decaying $\stone$ (see tables \ref{tab:stop-decay} 
and \ref{tab:fstates}). While all usual mass bounds obtained at colliders
for $\stone$ is generic for a stop-like state having the same
set of decay modes and hence applicable to the heavier stop as well,
a more practical and stringent sets of constraints respecting their 
hierarchy have recently been obtained by the LHC experiments
\cite{Aad:2014mha, Khachatryan:2014doa}.\footnote{An interesting caveat of 
these studies may be that these are silent 
about the decay branching fraction $\sttwo \to t \ntrlone$. This is 
unavoidable when they assume $\stone \to t \ntrlone$ is 100\%. It would be
interesting to see in what way BR[$\sttwo \to t \ntrlone$] affects these 
analyses.}
Whatever the case may be, these analyses are again expected to have
enhanced sensitivity thanks to the assumptions $\stone \to t \ntrlone$
is possible and has a 100\% branching fraction. This is consistent with the fact that more 
stringent bounds  are obtained from $\stone$ searches at the LHC with 
such an assumption when compared to those obtained in studies assuming 
$\stone \to c \ntrlone$ and $\stone \to b f \bar{f}^\prime \ntrlone$ as the 
only possibilities.

We now turn to the case of imminent LHC-13 followed by a possible case of
LHC-14 in the near future. In 
table \ref{tab:yields} we present yields for the backgrounds in various
different final states and the corresponding numbers for the signal events
accompanied by the signal significance (in parentheses) in each case for
an accumulated luminosity of 100 fb$^{-1}$.
These are done for all the three benchmark points and for both LHC-13 and
LHC-14. Kinematic cuts as presented in table \ref{tab:cuts} are used
in the analysis. The significances are estimated using the expression 
(see Appendix A of \cite{Ball:2007zza})
\[ \sigma = \sqrt{2 \left[ (S+B) \ln \left(1+{S\over B} \right) -S \right]} \]
which is applicable to situations with small number of events 
(in particular, when the number of background events is less than 50), 
is based on likelihood-ratios and follows from the Poisson distribution.

As can be seen from table \ref{tab:yields}, none of the modes except for the 
4-lepton final state ($\mathrm{SR4\ell 0b}$) could attain a $5\sigma$ 
signal-significance with 100 fb$^{-1}$ of data. It is further interesting to
note that for $\mathrm{SR4\ell 0b}$, a $\sim 5\sigma$ significance is
achievable for scenarios with a singlino-like LSP (BP3) or with one having a 
bino-like LSP (BP2) for which $\sbone$ mostly decays to $\stone W$, similar to 
as it does in the case of BP3. Note that the final state $\mathrm{SR3\ell 1b}$,
for the benchmark points BP2 and BP3, has the poorest strength of all reaching 
barely the $1\sigma$ level. This is not surprising since $\mathrm{SR3\ell 1b}$ 
requires a tagged $b$\emph{-jet} which, in our scenario, would mostly
come from 
the decay $\sbone \to b \ntrlone$. But this branching fraction is suppressed in 
BP2 and BP3 (see table \ref{tab:benchmarks}). Same is the explanation for a
poor show of the final state $\mathrm{SRSSDL1b}$ for BP2 and BP3.
In any case, it thus appears that these final states, which are somewhat of an 
exclusive kind as far as the number of jets (including the $b$-\emph{jets}) 
associated with each of them are concerned (see table \ref{tab:cuts}), are not 
very sensitive even to a moderately large volume of data at the imminent run(s) 
of the LHC.

The situation prompts us to explore an alternative strategy. It is found that 
looking for jet-inclusive final states for all these multi-lepton modes holds 
a better prospect. In table \ref{tab:yields-inclusive} we present the
corresponding numbers but after lifting the jet-related cuts (on multiplicity
and $p_T$). It is clear that a $\gtrsim 4 \sigma$ significance is possible
for final states with larger (3 to 4) lepton multiplicity. For the benchmarks
BP2 and BP3, signals in the inclusive 3- and 4-lepton final states could even attain
a significance of up to 6-7$\sigma$.  
It is also to be pointed out that as far as the
LHC phenomenology is concerned, the possibility of a large ($\sim 1$) 
$\mathrm{BR}\left[\sbone \to b \ntrlone \right]$ in the bino-dominated
LSP makes all the difference. This is the case with the benchmark point BP1.
It is clearly seen from table \ref{tab:yields} that the signal rates and their 
mutual patterns for BP1 are very different from those for BP2 and BP3. 
At the same time, the latter two, though represent different scenarios, 
have similar branching patterns for the sbottom and the stop squarks 
(see table \ref{tab:benchmarks}) which result in pretty similar rates and patterns for
various final states.
\begin{table}[!htbp]
\centering
  \begin{tabular}{ ||c|c||c|c|c|c|| }
    \hline \hline
 Channel &  $\sqrt{s}$ & \multicolumn{4}{c||}{Number of events and signal significances} \\
    \cline{3-6}
    ID   &   (in TeV)  & Background & BP1 ($\sigma$)& BP2 ($\sigma$) & BP3 ($\sigma$) \\
\hline \hline
 $\mathrm{SRSSDL0b}$    & $13$  &  9.7  &  1.8 (0.6)  &   7.3 (2.1)  &   8.6 (2.5) \\
    \cline{2-6}
                        & $14$  & 20.0  &  2.6 (0.6)  &  10.9 (2.3)  &  11.0 (2.3)    \\
\hline \hline
 $\mathrm{SRSSDL1b}$    & $13$  & 20.3  &  2.3 (0.5)  &   1.9 (0.4)  &   1.4 (0.3) \\
    \cline{2-6}
                        & $14$  & 26.1  &  2.4 (0.5)  &   2.4 (0.5)  &   1.9 (0.4)   \\
\hline \hline
 $\mathrm{SR3\ell0b}$  & $13$   & 12.1  &  4.1 (1.1)  &  11.3 (2.9)  &  11.5 (2.9)  \\
    \cline{2-6}
                       & $14$   & 20.0  &  5.1 (1.1)  &  15.6 (3.1)  &  15.5 (3.1)    \\
\hline \hline
 $\mathrm{SR3\ell 1b}$  & $13$  &  8.0  &  6.2 (2.0)  &   2.6 (0.9)  &   2.0 (0.7) \\
    \cline{2-6}
                        & $14$  &  5.4  &  7.9 (2.9)  &   3.4 (1.3)  &   2.4 (1.0)   \\
\hline \hline
 $\mathrm{SR4\ell0b}$  & $13$   &  1.7  &  5.4 (3.1)  &   8.7 (4.5)  &   9.8 (5.0) \\
    \cline{2-6}
                       & $14$   &  3.5  &  7.0 (3.0)  &  11.0 (4.4)  &  12.8 (4.9)     \\
\hline \hline
 \end{tabular}
 \caption{Number of background events in different final states and the same for
the signal events for the three benchmark points after cuts at LHC-13 and LHC-14. 
$\sigma$ in the parentheses corresponds to the significance 
level at an integrated luminosity of $100$ fb$^{-1}$.} 
\label{tab:yields}
\end{table} 
\begin{table}[!htbp]
\centering
  \begin{tabular}{ ||c|c||c|c|c|c|| }
    \hline \hline
Jet-inclusive & $\sqrt{s}$ & \multicolumn{4}{c||}{Number of events and signal significances} \\
    \cline{3-6}
final states       &   (in TeV)  & Background & BP1 ($\sigma$)& BP2 ($\sigma$) & BP3 ($\sigma$) \\
\hline \hline
 $\mathrm{SRSSDL}$  & $13$  & 175.9  &  18.2 (1.4)  &  28.5 (2.1)  &  29.6 (2.2) \\
    \cline{2-6}
                        & $14$  & 203.1  &  22.1 (1.5)  &  38.3 (2.6)  &  39.1 (2.7)   \\ 
\hline \hline
 $\mathrm{SR3\ell}$ & $13$  &  84.1  &  42.0 (4.3)  &  60.8 (6.0)  &  63.7 (6.3)  \\
    \cline{2-6}
                        & $14$  & 102.2  &  52.7 (4.8)  &  78.2 (7.0)  &  79.5 (7.1)    \\ 
\hline \hline
 $\mathrm{SR4\ell}$ & $13$  &  12.1  &  15.2 (3.7)  &  23.3 (5.4)  &  25.7 (5.9) \\
    \cline{2-6}
                        & $14$  &  21.0  &  20.5 (3.9)  &  30.6 (5.6)  &  33.3 (6.0)     \\ 
\hline \hline
   
 \end{tabular}
 \caption{Same as in table \ref{tab:yields} but for jet-inclusive final states.
See text for details.}
\label{tab:yields-inclusive}
\end{table} 
%
%
\begin{table} [!htbp]
\centering
\begin{tabular}{ |c|c|c|c| }
\hline
   Jet-inclusive &  \multicolumn{3}{c|}{Target Luminosity (fb$^{-1}$)}\\
    \cline{2-4}
   final states & BP1 & BP2 & BP3  \\ 
\hline
           $\mathrm{SRSSDL}$ &  1000 & 350 &  350 \\
           $\mathrm{SR3\ell}$ &  100 &  50 &  50 \\
           $\mathrm{SR4\ell}$ &  150 & 80 & 70 \\
\hline
\end{tabular}
\caption{Ballpark target (integrated) luminosities (in fb$^{-1}$) required at 
LHC-13/14 to obtain a 5$\sigma$ signal significance in various final states and 
for different benchmark points.
}
\label{tab:target}
\end{table} 
Furthermore, the usual expectation is that the significance for new physics
signal with rather massive states would go up as the collision energy grows.
In other words, the reach in mass is expected to increase with increasing 
energy of the experiment. Table \ref{tab:yields-inclusive} indeed shows such
an increase in the significance (for all the final states) with increasing
energy. However, the numbers show that LHC-13 could be already efficient
enough and simply pushing the machine energy to 14 TeV may not 
be particularly helpful, at least in the present context.

The overall sensitivity and/or reach of these experiments to the scenario in 
context are best demonstrated by the projected target (integrated) luminosities. 
In table \ref{tab:target} we present the required integrated luminosities 
(in fb$^{-1}$) that are needed to probe the three benchmark scenarios at 
different signal regions. 
As discussed in the last paragraph, given that they are expected to be rather 
similar for LHC-13 and LHC-14, we only present a single ballpark 
value of the target luminosity for each case.
The results can be summarized by saying that the simplified
scenario we propose with two light stops are most sensitive to final
states with large (3 to 4) lepton multiplicity. The luminosity requirements
for such modes in BP2 and BP3 reflect that these benchmark spectra copiously 
contribute to such multi-lepton final states.
%
%
\section{Conclusions}
%
A light stop is indispensable for a `natural' SUSY solution to the notorious 
hierarchy problem. Within the MSSM such a possibility is now under some tension 
as the observed value of the Higgs mass generically implies somewhat heavy stops. 
Set against this, we study the viability of not only one but both stops being 
relatively light (thus helping in resurrecting the `natural' SUSY) within the framework 
of vanilla ($Z_3$-symmetric) NMSSM and their implications for the LHC. The 
`$\lambda$' parameter of the NMSSM, associated with a new interaction term in 
the superpotential involving a singlet chiral superfield, gives rise to a 
tree-level contribution to the Higgs mass. This efficiently makes up for the 
lightness of the stops in such a scenario and liberates the stop sector from the 
Higgs sector to a pretty good extent.  We point out that direct searches at the 
LHC still allow for such light stops. 

We adopt a `minimal' scenario with a pair of light stops. A light sbottom is
then naturally present. The lighter stop is only a few tens of a GeV heavier 
than the LSP neutralino.  Thus, the only allowed decays of $\stone$ are its 
flavor-violating two-body decay to $c \ntrlone$ and its four-body decay to
$b f \bar{f}^\prime \ntrlone$. The LSP is also heavy enough to prohibit the 
decay $\sttwo \to t \ntrlone$. The decays $\sttwo \to \stone Z$ and/or 
$\sttwo \to \sbone \wplus$ are considered to be characteristic of the scenario. 
The benchmark scenarios studied in the present work have the mass-hierarchy 
$\mntrlone < \mstone < \msbone < \msttwo$. These low-lying states have their 
masses in the range $\sim$(300-500) GeV. Clearly, the spectra are of a 
`compressed' kind, with all other SUSY states taken to be heavy enough to get 
effectively decoupled. We point out that the compositions (singlino/bino 
contents) of the LSP neutralino could crucially influence the phenomenology of 
such a scenario.

The viability of such a scenario is demonstrated by subjecting the NMSSM 
parameter space to all relevant constraints from low energy electroweak data in 
various different sectors, to the latest constraints from the dark matter sector
and on the theoretical side, to the condition of absence of Landau poles.
Benchmark points are chosen based on the compositions of the LSP and the key 
decay modes of $\sttwo$ and $\sbone$. At colliders, such a scenario would lead 
to leptons, jets and $\slashed{E}_T$ all of which can at best be moderately 
hard. A priori, this may erode the sensitivity of the LHC experiments to the
scenario under consideration. 

Possible cascades of the stop and the sbottom squarks are discussed. $\stone$ 
decays promptly for the chosen benchmarks. Fast simulations (by including 
detector effects) are performed. Lepton-rich ($\geq 3$) events, with some of 
the lepton-pairs reconstructible to $Z$-bosons, are found to be rather 
characteristic of such a scenario. Finding a $b$-jet could be corroborative 
while tagging of $c$-jets to real benefit appears a bit futuristic.
Interestingly, even with three light squarks from the third generation, 
contrary to common expectations, some prominent final states could well have 
depleted or no $b$\emph{-jets}. SM backgrounds are simulated thoroughly.

It is found that the LHC-8 is not sensitive to the proposed scenario. The study 
shows that the LHC-13/14 with an accumulated luminosity of a few tens to a few 
hundreds of an inverse femtobarn would help explore the scenario. In the presence 
of signals, a multi-channel analysis could offer a quicker and holistic 
understanding of such a scenario. Furthermore, allowing for a hard ISR jet could 
help achieve harder final state objects and hence a cleaner signal region. The 
bottom-line is that such a scenario with two light stops may initially show up 
in very general final states which are not so typical in the search for stop 
squark(s). 

Before we close, the following few observations would be in place: 
\begin{itemize}
\item Existing and any future bounds on the stop mass assuming that it decays 
      only to flavor-violating $c\ntrlone$ and four-body 
      $b f \bar{f}^\prime \ntrlone$ would broadly be applicable to the scenario 
      we consider, irrespective of whether the LSP ($\ntrlone$) is bino- or 
      singlino-like, as long as the decays remain prompt. The same is true for 
      the sbottom if its decay to $b \ntrlone$ is 100\%. An eventual improvement 
      in these bounds, thus, would `slide' the allowed spectrum for such a 
      scenario to higher mass-values. When the decays are non-prompt, the bounds 
      from LHC-8 are generally rather severe, already touching nearly a TeV.
\item Depending on the relative values of $\mu_{eff}$, $M_2$ and $M_1$, there 
      may be situations when the lighter chargino and some other neutralinos 
      have much varied compositions and can become light enough to appear in 
      the cascades of the stops and the sbottoms. The combined LHC-phenomenology 
      of such light stop and sbottom squarks may then be rather involved. Its 
      study, however, is beyond the scope of the present work.
\item The possibility of a nearly degenerate pair of relatively light stops is
      also demonstrated. This would strip the scenario of the interesting 
      multi-lepton final states. On the top of that, this would alter the bounds 
      on the stop mass since, just like $\stone$, $\sttwo$ would also cascade to 
      $c\ntrlone$ and/or to $b f \bar{f}^\prime \ntrlone$ via an on-shell 
      $\stone$ and a highly virtual $Z$-boson that would lead to rather soft 
      fermion pairs escaping detection.
\item We consider a rather heavy ($\gtrsim 3$ TeV) gluino for our simplified, 
      minimal setup. Enhanced contributions to the inclusive final states are 
      foreseen for a relatively light gluino such that $\sttwo$-s could be
      copiously produced in its cascade. However, such yields are to be 
      subjected to the relevant LHC constraints which, in turn, would constrain 
      the scenario.
\item We also note that the reported  1-2 $\sigma$ level disagreement in the 
      measured $W^+W^-$ cross-section \cite{ATLAS:2012mec,Chatrchyan:2013yaa,Chatrchyan:2013oev} 
      with the SM prediction cannot be reproduced with our benchmark points. 
      The same is true for the $\sim 3 \sigma$ dilepton excess reported by ATLAS on-Z search \cite{Aad:2015wqa}. 
\end{itemize}

To summarize, a relatively light $\sttwo$ along with an even lighter $\stone$ 
is theoretically rather well-motivated but finding them together could be tricky.
In the NMSSM, the situation leads to interesting possibilities at the imminent 
runs of the LHC experiments by having these two states active in a rather 
collective way. However, such a possibility is rather generic to various other 
new physics scenarios with extra, possibly large contributions to the mass of 
the SM-like Higgs boson. Thus, a dedicated strategy to find such states is 
called for.
%

\acknowledgments
%
JB, AC and SKR are partially supported by funding available from the Department 
of Atomic Energy,  Government of India for the Regional Centre for 
Accelerator-based Particle Physics (RECAPP), Harish-Chandra Research Institute 
(HRI). The authors acknowledge the use of the cluster computing setup available 
at RECAPP and at the High Performance Computing facility of HRI. The authors 
would like to thank Debottam Das, Benjamin Fuks, Manoj Kumar Mandal, 
Olivier Mattelear, Florian Staub and Jamie Tattersall for very helpful 
discussions and Amit Khulve and Ravindra Yadav for technical support.
%
%
%

%
\end{document}